\documentclass[journal, onecolumn, 12pt]{IEEEtran}
\renewcommand{\baselinestretch}{2.0}
\usepackage{cite}
\usepackage{ifpdf}
\usepackage{array}
\ifpdf
\usepackage{graphicx}
\usepackage{subfigure}
\usepackage[svgnames]{xcolor}
\else
\usepackage[dvipdfmx]{graphicx}
\usepackage[dvipdfmx,svgnames]{xcolor}
\fi
\usepackage{amsmath,amssymb,amscd}
\usepackage{verbatim}
\usepackage{enumerate}
\usepackage{algorithm}
\usepackage{algpseudocode}
\usepackage{tikz}
\begin{document}
\renewcommand{\baselinestretch}{1.55}
\title{Cloud Radio Access Networks: Uplink Channel Estimation and Downlink Precoding}

\author{\large Osvaldo Simeone, Jinkyu Kang, Joonhyuk Kang and Shlomo Shamai (Shitz)
\thanks{O. Simeone is with the Center for Wireless Information Processing (CWIP), ECE Department, New Jersey Institute of Technology (NJIT), Newark, NJ 07102, USA (Email: osvaldo.simeone@njit.edu). 

Jinkyu Kang is with the School of Engineering and Applied Sciences (SEAS), Harvard University, Cambridge, MA 02138, USA (Email: jkkang@g.harvard.edu).

Joonhyuk Kang is with the Department of Electrical Engineering, Korea Advanced Institute of Science and Technology (KAIST) Daejeon, South Korea (Email: jhkang@ee.kaist.ac.kr).

S. Shamai (Shitz) is with the Department of Electrical Engineering, Technion, Haifa, 32000, Israel (Email: sshlomo@ee.technion.ac.il).
}
}
\maketitle
%
\section{Introduction}

The gains afforded by cloud radio access network (C-RAN) in terms of savings in capital and operating expenses, flexibility, interference management and network densification rely on the presence of high-capacity low-latency fronthaul connectivity between remote radio heads (RRHs) and baseband unit (BBU). In light of the non-uniform and limited availability of fiber optics cables, the bandwidth constraints on the fronthaul network call, on the one hand, for the development of advanced baseband compression strategies and, on the other hand, for a closer investigation of the optimal functional split between RRHs and BBU. In this chapter, after a brief introduction to signal processing challenges in C-RAN, this optimal function split is studied at the physical (PHY) layer as it pertains to two key baseband signal processing steps, namely channel estimation in the uplink and channel encoding/ linear precoding in the downlink. Joint optimization of baseband fronthaul compression and of baseband signal processing is tackled under different PHY functional splits, whereby uplink channel estimation and downlink channel encoding/ linear precoding are carried out either at the RRHs or at the BBU. The analysis, based on information-theoretical arguments, and numerical results yields insight into the configurations of network architecture and fronthaul capacities in which different functional splits are advantageous. The treatment also emphasizes the versatility of deterministic and stochastic successive convex approximation strategies for the optimization of C-RANs.

\section{Technology Background}
In a C-RAN architecture, the base station (BS) functionalities, from the PHY layer to higher layers, are implemented in a virtualized fashion on centralized general-purpose processors rather than on the local hardware of the base stations or access points. This results in a novel cellular architecture in which low-cost wireless access points$-$the RRHs$-$which retain only radio functionalities, are centrally managed by a reconfigurable centralized ``cloud", the BBU. At a high level, the C-RAN concept can be seen as an instance of network function virtualization and hence as the RAN counterpart of the separation of control and data planes proposed for the core network in software-defined networking \cite{Han}.

The C-RAN architecture has the following key advantages, which make it a key contender for inclusion in a 5G standard:

\begin{itemize}

    \item Reduced capital expense due to the possibility to substitute full-fledged base stations with RRHs with reduced space and energy requirements;
    \item Statistical multiplexing gain thanks to the flexible allocation of radio and computing resources across all the connected RRHs;
    \item Easier implementation of coordinated and cooperative transmission/ reception strategies, such as Enhanced Inter-Cell Interference Coordination (eICIC) and Coordinated MultiPoint (CoMP) in Long Term Evolution Advanced (LTE-A), to mitigate multi-cell interference;
    \item Simplified network upgrades and maintenance owing to the centralization of RAN functionalities.

\end{itemize}

The C-RAN architecture depends on a network of so-called fronthaul links to enable the virtualization of BS functionalities at a BBU. This is because in the uplink, the RRHs are required to convey their respective received signals, either in analog format or in the form of digitized baseband samples, to the BBU for processing. Moreover, in a dual fashion, in a C-RAN downlink, each RRH needs to receive from the BBU either directly the analog radio signal to be transmitted on the radio interface, or a digitized version of the corresponding baseband samples. The RRH$-$BBU bidirectional links that carry such information are referred to as \textit{fronthaul} links, in contrast to the backhaul links connecting the BBU to the core network.

The analog transport solution is typically implemented on fronthaul links by means of radio-over-fiber \cite{RoF}. Instead, the digital transmission of baseband, or IQ, samples is currently carried out by following the Common Public Radio Interface (CPRI) standard \cite{CPRI}, which most commonly requires fiber optic fronthaul links as well. The digital approach appears to be favored due to the traditional advantages of digital solutions, including resilience to noise and hardware impairments and flexibility in the transport options \cite{Checko}.

\subsection{Signal Processing Challenges in C-RAN}

The main roadblock to the realization of the mentioned promises of C-RAN hinges on the inherent restrictions on bandwidth and latency of the fronthaul links that may limit the advantages of centralized processing at the BBU. 

\subsubsection{Fronthaul capacity limitations} Implementing the CPRI standard, the bit rate required for base station that serve multiple cell sectors with carrier aggregation and with multiple antennas exceeds the 10 Gbit/s provided by standard fiber optics links \cite{Checko}, \cite{IDT}. This problem is even more pronounced for networks in which fiber-optic links are not available due to the large expense required for their deployment or lease, as for heterogeneous networks with smaller RRHs \cite{Fujitsu}. 
The capacity limitations of the fronthaul link call for the development of compression strategies that reduce the fronthaul rate with minor or no degradation in the quality of the quantized baseband signal. Typical solutions are based on filtering, per-block scaling, lossless compression, predictive quantization, see \cite{Samardzija, Guo, Nieman13GlobalSIP, Lorca, Grieger, Vosoughi} 

When quantization and compression are not sufficient, as reported in \cite{Dotsch}, \cite{Wubben}, the bottleneck on the performance of C-RANs due to the capacity limitations of the fronthaul links can be alleviated by implementing a more flexible separation of functionalities between RRHs and BBU, rather than performing all baseband processing at the BBU. Examples of baseband operations that can be carried out at the RRH include Fast Fourier Transform and Inverse Fast Fourier Transform (FFT and IFFT), demapping, synchronization, channel estimation, precoding and channel encoding. Note that \cite{Dotsch} also investigates the possibility to implement functions at higher layers, such as error detection, at the RRHs. We will elaborate on important aspects of the functional split between RRH and BBU below.
    
\subsubsection{Fronthaul latency limitations} Two of the communication protocols that are most affected by fronthaul delays are uplink hybrid automatic repeat request (HARQ) and random access \cite{Dotsch}. For HARQ, the problem is that the outcome of decoding at the BBU may only become available at the RRH after the time required for 

\begin{itemize}

\item the transfer of the baseband signals from the RRH to the BBU
\item the processing at the BBU 
\item the transmission of the decoding outcome from the BBU to the RRH. 

\end{itemize}

This delay may seriously affect the throughput achievable by uplink HARQ. For example, in LTE with frequency division multiplexing, the feedback latency should be less than $8$ ms in order not to disrupt the operation of the system \cite{Dotsch}. Similar issues impair the implementation of random access.

\subsection{Chapter Overview}
In this chapter, we explore the problem of optimal functional split between RRHs and BBU at the PHY layer by focusing on the two key baseband operations of channel encoding and channel encoding/ precoding. We recall that alternative functional splits are envisaged to be potentially advantageous in the presence of significant fronthaul capacity constraints. 

For the uplink, we compare the standard implementation in which all baseband processing, including channel estimation, is performed at the BBU, with an alternative architecture in which channel estimation, along with the necessary frame synchronization and resource demapping, is instead implemented at the RRHs. This is discussed in Sec. \ref{sec:uplink}.

The downlink is discussed in Sec. \ref{sec:downlink}, where we contrast the standard C-RAN implementation with an alternative one in which channel encoding and precoding are applied at the RRHs, while the BBU retains the function of designing the precoding matrices based on the available channel state information.

Throughout, we take an information-theoretic approach in order to evaluate analytical expressions for the achievable performance that illuminates the impact of different design choices. The analysis is corroborated by extensive numerical results that provide insight into the performance comparisons highlighted above. The chapter is concluded in Sec. \ref{sec:Conclusion}.

\section{Uplink: Where to Perform Channel Estimation?}\label{sec:uplink}

In this section, we study the uplink and address the potential advantages that could be accrued by performing channel estimation at the RRHs rather than at the BBU. The rationale for the exploration of this functional split is that communicating the digitized signal received within the training portion of the received signal, as done in the conventional implementation, may impose a more significant burden on the fronthaul network that communicating directly the estimated channel state information (CSI). This split is also supported by the known information-theoretic optimality of separate estimation and compression \cite{Witsenhausen80TIT}. In particular, we compare two different approaches:

\begin{itemize}

\item the conventional approach, in which the RRHs quantize the training signals and CSI estimation takes place at the BBU; 
\item channel Estimation at the RRHs, in which the RRHs perform CSI estimation and forward a quantized version of the CSI to the BBU. 
 
\end{itemize}
Note that the conventional approach was the subject of an earlier study \cite{Kobayashi11TSP} and that this section is adapted from our earlier work \cite{Kang14TWC}, to which we refer for proofs and additional considerations.

We start by discussing the system model in Sec. \ref{sec:system model uplink} and then elaborate on the two approaches in Sec. \ref{subSec:CFE_Pre} and Sec. \ref{sec:CE RRH}. Finally, we present numerical results in Sec. \ref{Sec:Numerical Results uplink}.

\label{Sec:SM}
\begin{figure}[t]
\centering
\includegraphics[width=13.5cm]{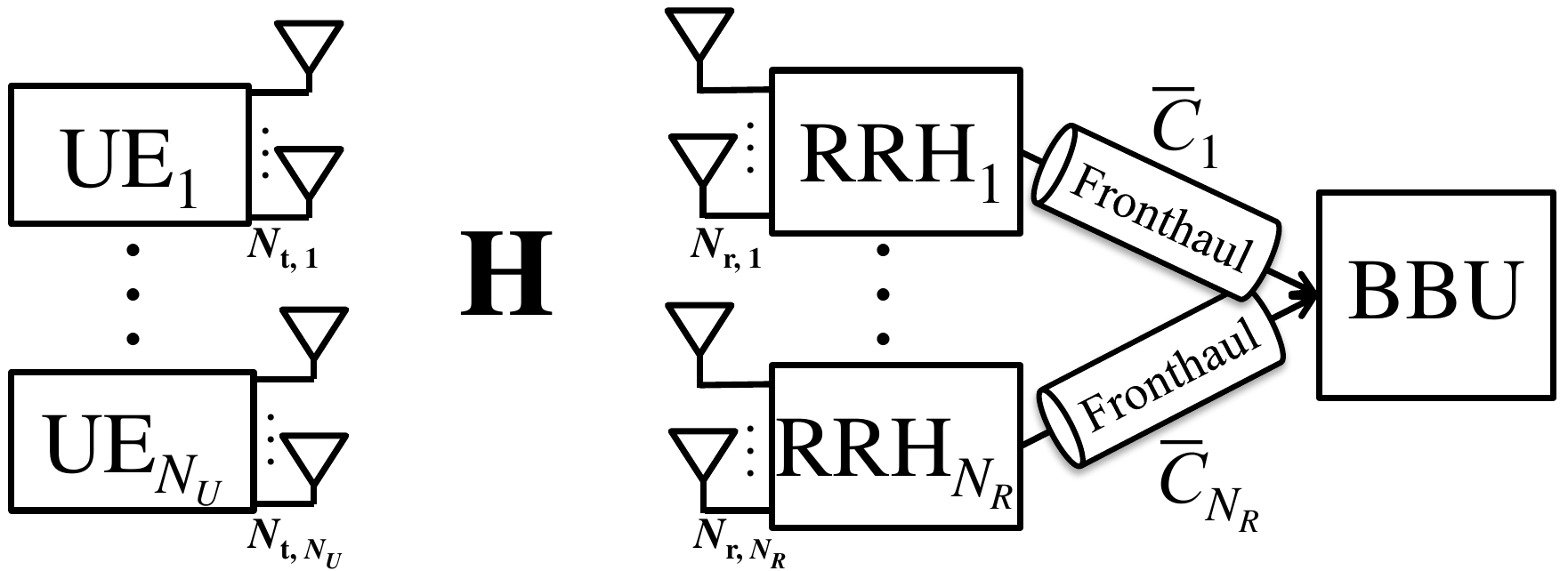}
\caption{Uplink of a C-RAN system consisting of $N_U$ UEs and $N_R$ RRHs. Each $j$-th RRH is connected to the BBU with a fronthaul link of capacity $\bar C_j$.}
\label{fig:fig1}
\end{figure}

\subsection{System Model}\label{sec:system model uplink}

We study the uplink of a cellular system consisting of $N_U$ User Equipments (UEs), $N_R$ RRHs and a BBU, as shown in Fig. \ref{fig:fig1}. We denote the set of all UEs, or mobile users, as $\mathcal{N}_U = \{1, \dots, N_U\}$ and the set of all RRHs as $\mathcal{N}_R = \{1, \dots, N_R\}$. Each $i$-th UE has $N_{t,i}$ transmit antennas, while each $j$-th RRH is equipped with $N_{r,j}$ receive antennas. We define the number of total transmit antennas as $N_t = \sum_{i=1}^{N_U} N_{t,i}$. Each $j$-th RRH is connected to the BBU via a fronthaul link of capacity $\bar C_j$. All rates, including $\bar C_j$, are normalized to the bandwidth available on the uplink channel from the UEs to the RRHs and are measured in bits/s/Hz. We assume that coding is performed across a large number of channel coherence blocks, for example over many resource blocks of an LTE system operating on a channel with significant time-frequency diversity. This implies that the ergodic capacity describes the system performance in terms of achievable rates (see, e.g., \cite{Hassibi03TIT}).

Each channel coherence block, of length $T$ channel uses, is split into a phase for channel training of length $T_p$ channel uses and a phase for data transmission of length $T_d$ channel uses, with
\begin{equation}
T_p + T_d = T. \label{TC}
\end{equation}
The signal transmitted by the $i$-th UE is given by a $N_{t,i} \times T$ complex matrix ${\bf{X}}_i$, where each column corresponds to the signal transmitted by the $N_{t,i}$ antennas in a channel use. This signal is divided into the $N_{t,i} \times T_p$ pilot signal ${\bf{X}}_{p,i}$ and the $N_{t,i} \times T_d$ data signal ${\bf{X}}_{d,i}$. We assume that the transmit signal ${\bf{X}}_i$ has a total per-block power constraint $T^{-1}E[\| {\bf{X}}_i \|^2] = \bar P_i$, and we define ${T_p}^{-1}E[\| {\bf{X}}_{p,i} \|^2] = P_{p,i}$ and ${T_d}^{-1} E[\| {\bf{X}}_{d,i} \|^2]  = P_{d,i}$ as the powers used for training and data, respectively by the $i$-th UE. Note that $E[ \cdot ]$ refers throughout to the expectation operator. In terms of pilot and data signal powers, the power constraint is hence expressed as
\begin{equation}
 \frac{T_p }{T}P_{p,i}  + \frac{T_d }{T} P_{d,i}  = \bar P_i. \label{PC}
\end{equation}
For simplicity, we assume equal transmit power allocation for all UEs, and hence we have $\bar P_i = \bar P$, $P_{d,i} = P_d$ and $P_{p,i}=P_p$ for all $i \in \mathcal{N}_U$. Finally, we collect in matrices ${\bf{X}}_p$ and ${\bf{X}}_d$ all the pilot signals and the data signals transmitted by all UEs, respectively, i.e., ${\bf{X}}_p = [{\bf{X}}_{p,1}^T, \dots, {\bf{X}}_{p,N_U}^T]^T$ and ${\bf{X}}_d = [{\bf{X}}_{d,1}^T, \dots, {\bf{X}}_{d,N_U}^T]^T$.

The training signal is ${\bf{X}}_p = {\sqrt{{P_p}/{N_t}}} {\bf{S}}_p$, where ${\bf{S}}_p$ is a $N_t \times T_p$ matrix with orthogonal rows and unitary power entries corresponding to the orthogonal training sequences transmitted from each antenna by all UEs (as in, e.g., \cite{Kobayashi11TSP}). Note that this implies that each training sequence is transmitted with power $P_p/N_t$ and that the condition $T_p \geq N_t$ holds. During the data phase, the UEs transmit independent space-time codewords without precoding. Using random coding arguments, we write ${\bf{X}}_d = {\sqrt{{P_d}/{N_t}}} {\bf{S}}_d$, where ${\bf{S}}_d$ is a $N_t \times T_d$ matrix of independent and identically distributed (i.i.d.) $\mathcal{CN}(0,1)$ variables.

The $N_{r,j} \times T$ signal ${\bf{Y}}_j$ received by the $j$-th RRH in a given coherence block, where each column corresponds to the signal received by the $N_{r,j}$ antennas in a channel use, can be split into the $N_{r,j} \times T_p$ received pilot signal ${\bf{Y}}_{p,j}$ and the $N_{r,j} \times T_d$ data signal ${\bf{Y}}_{d,j}$. The signal received at the $j$-th RRH is then given by
\begin{subequations}
\label{RS;RRH}
\begin{eqnarray}
\label{RS_p;RRH} {\bf{Y}}_{p,j} &=& \sqrt{\frac{P_p}{N_t}} {\bf{H}}_j {\bf{S}}_p  + {\bf{Z}}_{p,j} \\
\label{RS_d;RRH} \textrm{and}\,\,\,\, {\bf{Y}}_{d,j} &=& \sqrt{\frac{P_d}{N_t}} {\bf{H}}_j {\bf{S}}_d  + {\bf{Z}}_{d,j},
\end{eqnarray}
\end{subequations}
where ${\bf{Z}}_{p,j}$ and ${\bf{Z}}_{d,j}$ are respectively the $N_{r,j} \times T_p$ and $N_{r,j} \times T_d$ matrices of i.i.d. complex Gaussian noise variables with zero-mean and unit variance, i.e., $\mathcal{CN}(0,1)$. The $N_{r,j} \times N_t$ channel matrix ${\bf{H}}_j$ collects all the $N_{r,j} \times N_{t,i}$ channel matrices ${\bf{H}}_{ji}$ from the $i$-th UE to the $j$-th RRH as ${\bf{H}}_j = [{\bf{H}}_{j1}, \dots, {\bf{H}}_{jN_U}]$.

The channel matrix ${\bf{H}}_{ji}$ is modeled as having i.i.d. $\mathcal{CN}(0,\alpha_{ji})$ entries, where $\alpha_{ji}$ is the path loss coefficient between the $i$-th UE and the $j$-th RRH being given as
\begin{equation} \label{Chap3_PL_coef}
\alpha_{ji} = \frac{1}{1 + \left (\frac{d_{ji}}{d_0}\right )^{\eta}},
\end{equation}
where $d_{ji}$ is the distance between the $i$-th UE and the $j$-th RRH, $d_0$ is a reference distance, and $\eta$ is the path loss exponent. The channel matrices are assumed to be constant during each channel coherence block and to change according to an ergodic process from  block to block.
\subsection{Conventional Approach}
\label{subSec:CFE_Pre} With the conventional approach, the RRH quantizes and compresses both its received pilot signal in Eq. (\ref{RS_p;RRH}) and its received data signal in Eq. (\ref{RS_d;RRH}), and forwards the compressed signals to the BBU on the fronthaul link. The BBU then estimates the CSI on the basis of the received quantized pilot signals and performs coherent decoding of the data signal. In the rest of Sec. \ref{sec:uplink}, we limit the analytical treatment to the case of a single UE and a single RRH, i.e., $N_U=1$ and $N_R=1$, for simplicity of presentation. We henceforth remove the subscripts indicating UE and RRH indices. A more general discussion can be found elsewhere \cite{Kang14TWC}.

\subsubsection{Training Phase}
During the training phase, the vector of received training signals $ {\bf{Y}}_{p}$ in Eq. (\ref{RS_p;RRH}) across all coherence times is quantized. In order to account for quantization and compression, throughout this chapter, we use the standard additive quantization noise model that follows conventional information-theoretical arguments based on random coding \cite{GamalBook}. Accordingly, the quantized pilot signal can be written as
\begin{equation}
\widehat {\bf{Y}}_{p} = {\bf{Y}}_{p}  + {\bf{Q}}_{p}, \label{CTS;BBU}
\end{equation}
where the compression noise matrix ${\bf{Q}}_{p}$ is assumed to have i.i.d.  $\mathcal{CN} (0, \sigma_{p}^2)$ entries. Note that the assumption of Gaussian i.i.d. quantization noises is made here for simplicity of analysis without claim of optimality. On a practical note, Gaussian quantization noise can be realized by high-dimensional vector quantizers such as trellis-coded quantization \cite{Zamir}. The quantization noise variance $\sigma_{p}^2$ dictates the accuracy of the quantization and depends on the fronthaul capacity via standard information-theoretic identities \cite{GamalBook}, as further discussed below. \par
Based on Eq. (\ref{CTS;BBU}), the channel matrix ${\bf{H}}$ from the UE to the RRH is estimated at the BBU by the minimum mean square error (MMSE) method. Hence, it can be expressed as
\begin{equation}
{\bf{H}} = \widehat {\bf{H}} + {\bf{E}}, \label{ECS+EE;BBU}
\end{equation}
where the estimated channel $\widehat {\bf{H}}$ is a complex Gaussian matrix with i.i.d. $\mathcal{CN} (0, \sigma_{\widehat h}^2)$ entries, and the estimation error ${\bf{E}}$ has i.i.d. $\mathcal{CN} (0, \sigma_{e}^2)$ entries. With $\sigma_{\widehat h}^2 = \alpha - \sigma_{e}^2$ and $\sigma_{e}^2 = { \alpha N_t (1+\sigma_{p}^2)}/({T_p P_p +  N_t (1+\sigma_{p}^2)})$, respectively \cite{Hassibi03TIT}, \cite{Bjornson10TSP}, where we recall that $\alpha$ is the power gain for the channel between UE and RRH.

\subsubsection{Data Phase}
The quantized data signal received at the BBU can be similarly expressed as $\widehat {\bf{Y}}_{d} = {\bf{Y}}_{d}  + {\bf{Q}}_{d}$, where the quantization noise ${\bf{Q}}_{d}$ is assumed to have i.i.d.  $\mathcal{CN} (0, \sigma_{d}^2)$ entries. Moreover, it can be  written as the sum of a useful term $\widehat {\bf{H}} {\bf{X}}_d$ and of the equivalent noise ${\bf{N}}_{d} = {\bf{E}} {\bf{X}}_d + {\bf{Z}}_{d} + {\bf{Q}}_{d}$, namely
\begin{equation}
\widehat {\bf{Y}}_{d} = \widehat {\bf{H}} {\bf{X}}_d + {\bf{N}}_{d}, \label{RDS;BBU:ComForEst}
\end{equation}
where the equivalent noise ${\bf{N}}_{d}$ has i.i.d. entries with zero mean and power $1 + \sigma_{d}^2 + {P_d} \sigma_{e}^2$. We observe that ${\bf{N}}_d$ is not Gaussian distributed and is not independent of ${\bf{X}}_d$. Further discussion can be found in the literature \cite{Kang14TWC, Hassibi03TIT}.

\subsubsection{Ergodic Rate}
As mentioned, we adopt as the performance criterion of interest the ergodic rate, which, under the assumption of Gaussian codebooks, is given by the mutual information $T^{-1} I ( {\bf{X}}_d; \widehat {\bf{Y}}_d | \widehat {\bf{H}} )$ [bits/s/Hz] (see, e.g, \cite[Ch. 3]{GamalBook}). This quantity can be lower-bounded by the following expression \cite{Kang14TWC}:
\begin{eqnarray}
R = \frac{T_d}{T} E \left [ \log_2 \det \left( {\bf{I}}_{N_r} + \rho_{\textrm{eff}} \widehat {\bf{H}} \widehat {\bf{H}}^\dagger \right) \right], \label{EMI;ComForEst}
\end{eqnarray}
with $\rho_{\textrm{eff}} = {P_d}/({N_t (1 + \sigma_d^2 + {P_d} \sigma_{e}^2 )})$ being the effective signal to noise ratio (SNR), which accounts for the effects of quantization and channel estimation, and $\widehat {\bf{H}}$ being distributed as in Eq. (\ref{ECS+EE;BBU}). The rate in Eq. (\ref{EMI;ComForEst}) is hence an achievable ergodic rate \cite{Kang14TWC}. Moreover, let us define as $C_p$ the fronthaul rate allocated to transmit information about the pilot signals and as $C_d$ the fronthaul rate for the data with  $C_p + C_d = \bar C$. Then, if the conditions
\begin{subequations}
\label{BR;CFE}
\begin{eqnarray}
\label{BR;CFE_DS} C_p \hspace{-0.2cm} &=&  \hspace{-0.2cm} \frac{T_p N_r}{T} \log_2 \left( 1 + \frac{ P_p \alpha + 1 }{\sigma_{p}^2} \right) \\
\label{BR;CFE_TS} \textrm{and}  \hspace{0.2cm} C_d  \hspace{-0.2cm} &=&  \hspace{-0.2cm} \frac{T_d N_r}{T} \log_2 \left( 1 + \frac{ P_d \alpha + 1 }{\sigma_{d}^2} \right)
\end{eqnarray}
\end{subequations}
are satisfied, a quantization (and compression) scheme exists that guarantees the desired quantization errors $(\sigma_d^2, \sigma_p^2)$ \cite{Kang14TWC}.

The ergodic achievable rate in Eq. (\ref{EMI;ComForEst}) can now be optimized over the fronthaul allocation $( C_{p}, C_{d} )$ under the fronthaul constraint $\bar C=C_p+C_d$, with $C_p$ and $C_d$ in Eq. (\ref{BR;CFE}), by maximizing the effective SNR $\rho_{\textrm{eff}}$ in Eq. (\ref{EMI;ComForEst}). This non-convex problem can be tackled using a line search method \cite{BoydBook} in a bounded interval (e.g., over $C_p$ in the interval $[0, \bar C]$).

\subsection{Channel Estimation at the RRHs}\label{sec:CE RRH}
With the mentioned alternative functional split, each RRH estimates the CSI on the basis of its received pilot signal in Eq. (\ref{RS_p;RRH}), and then quantizes and compresses both its estimated CSI and its received data signal in Eq. (\ref{RS_d;RRH}) for transmission on the fronthaul.

\subsubsection{Training Phase}
The RRH performs the MMSE estimate of the channel ${\bf{H}}$ given the observation ${\bf{Y}}_{p}$ in Eq. (\ref{RS_p;RRH}). As a result, similar to Eq. (\ref{ECS+EE;BBU}), we can decompose the channel matrix ${\bf{H}}$ into the MMSE estimate $\widetilde {\bf{H}}$ and the independent estimation error ${\bf{E}}$, as
\begin{equation}
{\bf{H}} = \widetilde {\bf{H}} + {\bf{E}}, \label{ECS+EE}
\end{equation}
where the error ${\bf{E}}$ has i.i.d. $\mathcal{CN} (0, \sigma_{e}^2)$ entries with $\sigma_{e}^2 =  {\alpha N_t}/(T_p P_p + N_t)$ and $\widetilde {\bf{H}}$ has i.i.d. $\mathcal{CN} (0, \sigma_{\widetilde h}^2)$ entries with $\sigma_{\widetilde h}^2 = \alpha - \sigma_e^2$.

The sequence of channel estimates $\widetilde {\bf{H}}$ for all coherence times in the coding block is compressed by the RRH and forwarded to the BBU on the fronthaul link. The compressed channel $\widehat {\bf{H}}$ is related to the estimate $\widetilde {\bf{H}}$ as
\begin{equation}
\widetilde {\bf{H}} = \widehat {\bf{H}} + {\bf{Q}}_{p}, \label{CCS+QN}
\end{equation}
where the $N_{r} \times N_t$ quantization noise matrix ${\bf{Q}}_{p}$ has i.i.d. $\mathcal{CN}(0,\sigma_{p}^2)$ entries.
\subsubsection{Data Phase}
\label{Sec:DP;SM} During the data phase, the RRH quantizes the signal ${\bf{Y}}_{d}$ in Eq. (\ref{RS_d;RRH}) and sends it to the BBU on the fronthaul link. The signal obtained at the BBU is related to ${\bf{Y}}_{d}$ as
\begin{equation}
\widehat {\bf{Y}}_{d} = {\bf{Y}}_{d}  + {\bf{Q}}_{d}, \label{RDS;RRH}
\end{equation}
where ${\bf{Q}}_{d}$ is independent of ${\bf{Y}}_{d}$ and represents the quantization noise matrix with i.i.d. $\mathcal{CN}(0,\sigma_{d}^2)$ entries. Separating the desired signal and the noise in Eq. (\ref{RDS;RRH}), the quantized signal $\widehat {\bf{Y}}_{d}$ can be expressed as
\begin{eqnarray}
\widehat {\bf{Y}}_{d} = \widehat {\bf{H}} {\bf{X}}_d + {\bf{N}}_{d}, \label{RDS;BBU}
\end{eqnarray}
where ${\bf{N}}_{d}$ denotes the equivalent noise ${\bf{N}}_{d} = \left( {\bf{Q}}_{p} + {\bf{E}} \right) {\bf{X}}_d + {\bf{Z}}_{d} + {\bf{Q}}_{d}$, which has i.i.d. zero-mean entries with power
\begin{equation}
\label{Noise_pe} \sigma_{n}^2 = P_d \left( \sigma_{p}^2 + \sigma_{e}^2 \right) 1 + \sigma_d^2.
\end{equation}
We observe that, as in Eq. (\ref{RDS;BBU:ComForEst}), ${\bf{N}}_{d}$ is not Gaussian distributed and is not independent of ${\bf{X}}_d$.

\subsubsection{Ergodic Rate}
Let $C_p$ and $C_d$ denote respectively the fronthaul rates allocated for the transmission of the quantized channel estimates in Eq. (\ref{CCS+QN}) and of the quantized received signals in Eq. (\ref{RDS;RRH}) on the fronthaul link from the RRH to the BBU. An achievable ergodic rate is given as \cite{Kang14TWC}:
\begin{eqnarray}
R = \frac{T_d}{T} E \left[ \log _2 \det \left( {\bf{I}}_{N_r} + \rho_{{\textrm{eff}}} { \widehat {\bf{H}}} {\widehat {\bf{H}}}^\dagger \right) \right], \label{EAR;SC_IC}
\end{eqnarray}
with the effective SNR
\begin{equation}
\rho_{{\textrm{eff}}}  = \frac{P_d}{N_t \sigma_n^2} = \frac{P_d}{ N_t \left( 1 + \sigma_{d}^2 + P_d \left( \sigma_{p}^2 +  \sigma_{e}^2 \right) \right) }; \label{ESNR;SC_IC}
\end{equation}
$\widehat {\bf{H}}$ being distributed as in Eq. (\ref{CCS+QN}); and with $\sigma_{e}^2$ in Eq. (\ref{ECS+EE}). Moreover, if the conditions
\begin{subequations}
\label{BR;ECF_Sep}
\begin{eqnarray}
\label{BR;ECF_Sep_DS} C_p \hspace{-0.2cm}&=& \hspace{-0.2cm}\frac{N_r N_t}{T} \log_2 \left( \frac{ \alpha - \sigma_{e}^2 }{\sigma_{p}^2} \right) \\
\label{BR;ECF_Sep_TS} \textrm{and} \hspace{0.3cm} C_d \hspace{-0.2cm} &=& \hspace{-0.2cm} \frac{N_r T_d}{T} \log_2 \left( 1 + \left( \frac{\alpha P_d + 1}{\sigma_d^2} \right) \right),
\end{eqnarray}
\end{subequations}
are satisfied, then a quantization scheme exists that guarantees the desired quantization error $(\sigma_p^2, \sigma_d^2)$ \cite{Kang14TWC}. The ergodic achievable rate in Eq. (\ref{EAR;SC_IC}) can now be optimized over the fronthaul allocation $( C_{p}, C_{d} )$ under the fronthaul constraint $\bar C=C_p+C_d$, with $C_p$ and $C_d$ in Eq. (\ref{BR;ECF_Sep}), by maximizing the effective SNR $\rho_{\textrm{eff}}$ in Eq. (\ref{ESNR;SC_IC}) using a line search \cite{BoydBook} in a bounded interval.

\subsubsection{Adaptive quantization}
The alternative functional split studied here enables the RRHs to performs adaptive quantization of the data as a function of the estimated CSI in each coherence block. Specifically, rather than performing separate quantization of CSI and data, the data is quantized in each coherence period with a different accuracy depending on the corresponding CSI: a better channel quality calls for a more accurate quantization of the data field, and vice versa for worse CSI. We note that this is not possible in the conventional approach in which CSI is not estimated at the RRHs. Further details can be found elsewhere \cite{Kang14TWC}.

\subsection{Numerical Results}
\label{Sec:Numerical Results uplink}

\begin{figure}[t]
\centering
\includegraphics[width=11cm]{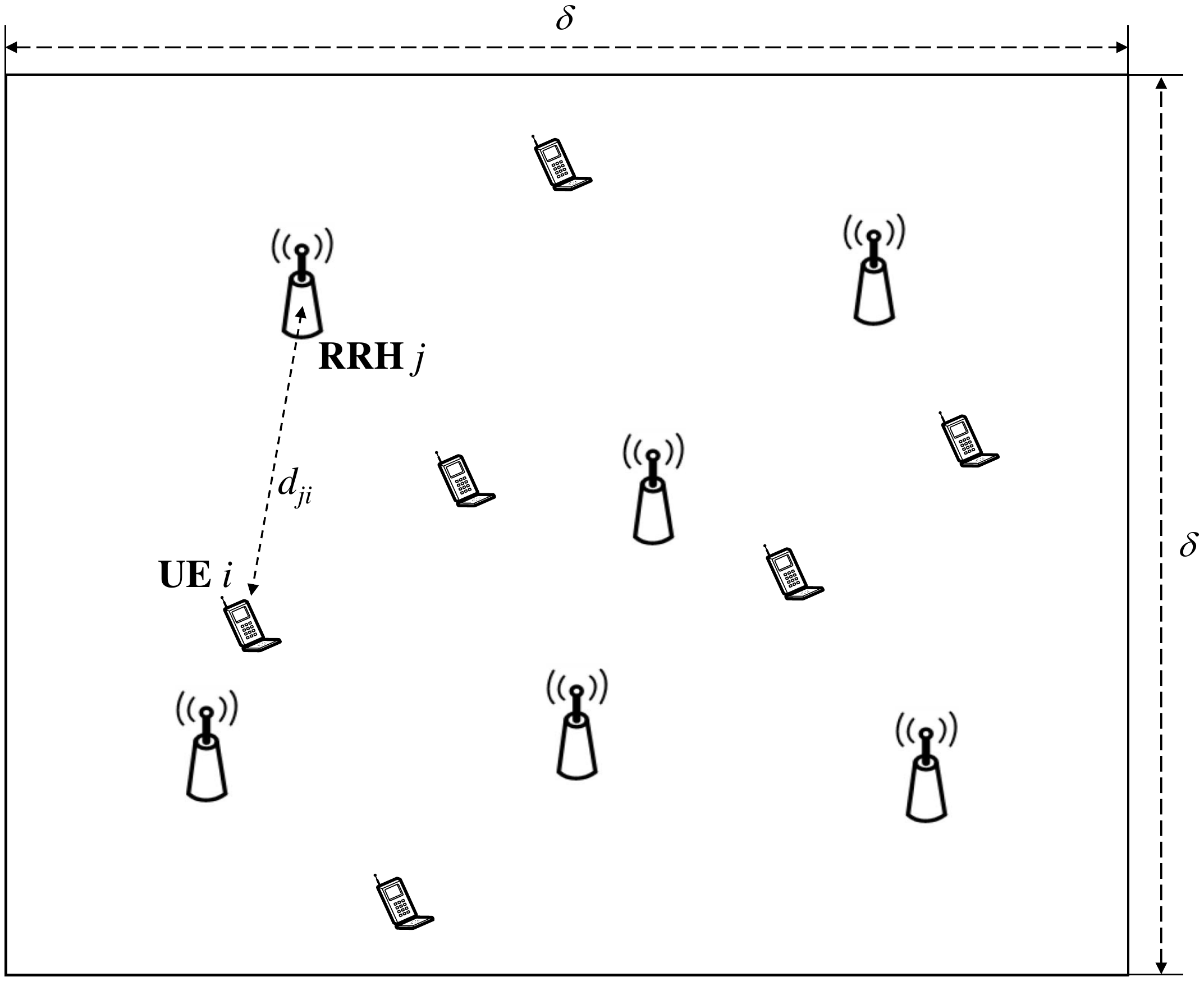}
\caption{Set-up under consideration for the numerical results, where RRHs and UEs are located in a square with side $\delta$. All RRHs are connected to the same BBU.}
\label{Chap3_Fig:SimulEnvironment}
\end{figure}

\begin{figure}[t]
\centering
\includegraphics[width=13.5cm]{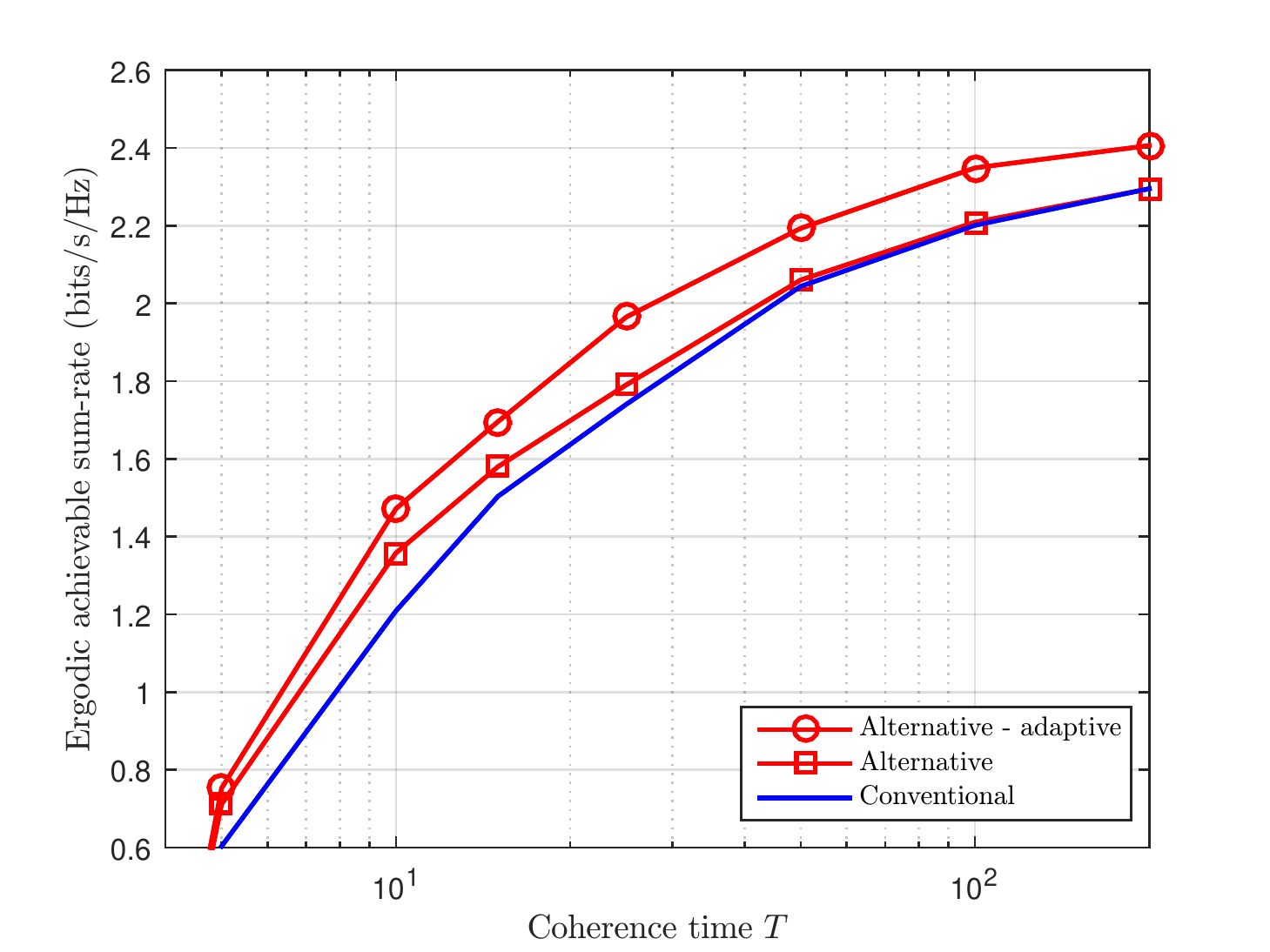}
\caption{Ergodic achievable sum-rate vs. coherence time ($N_R = N_U = 2$, $N_t = N_r=4$, $\bar C$ = 6 bits/s/Hz, and $\bar P=10dB$).}
\label{Fig:UplinkvsCT}
\end{figure}

\begin{figure}[h!]
\centering
\includegraphics[width=13.5cm]{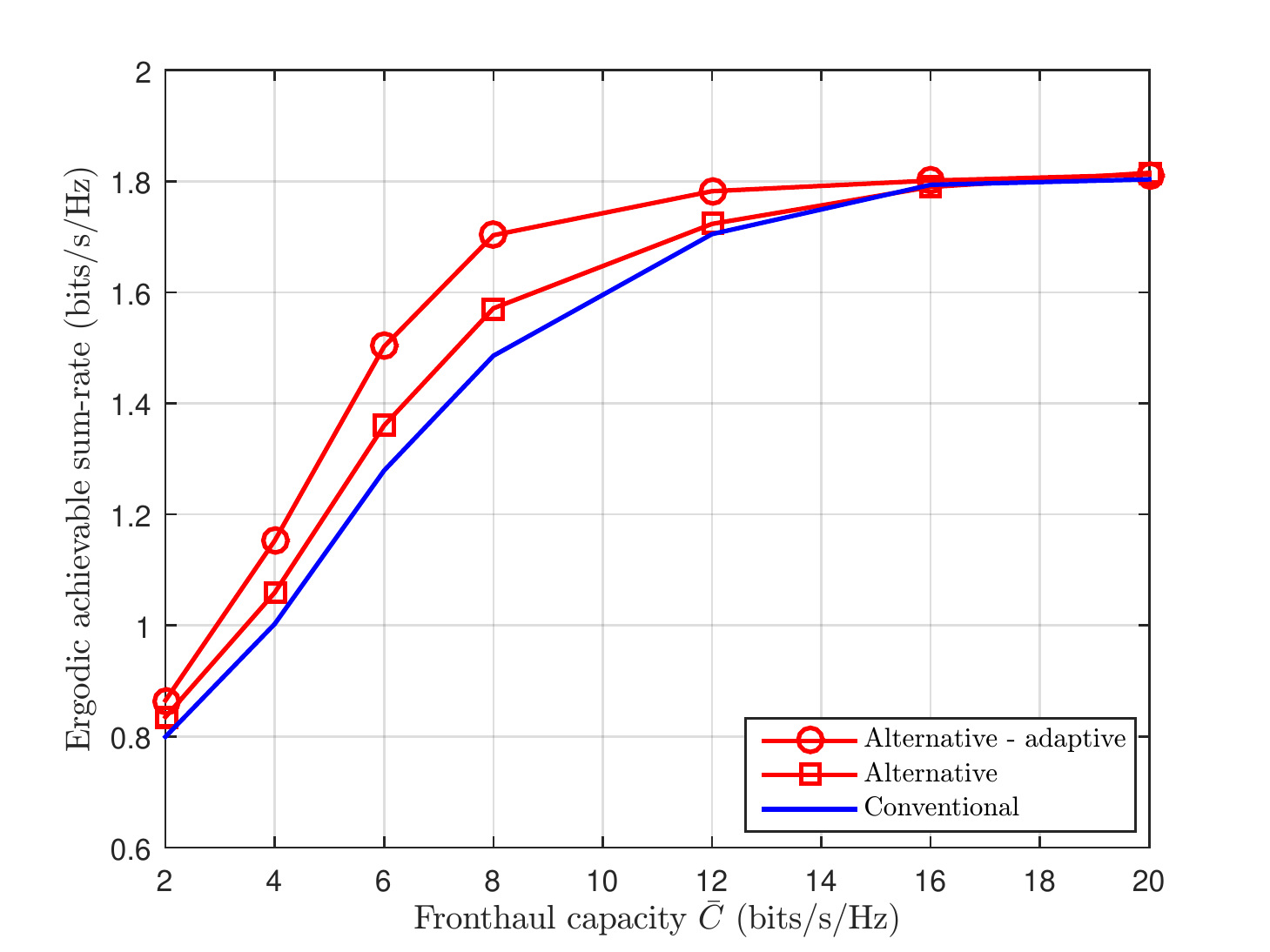}
\caption{Ergodic achievable sum-rate vs. fronthaul capacity ($N_R = N_U = 2$, $N_t=N_r=4$, $\bar P=10dB$, and $T$ = 10).}
\label{Fig:UplinkvsFC}
\end{figure}

In this section, we evaluate the performance of the discussed conventional and alternative strategies for the uplink. For the latter, we consider both the basic and adaptive implementations mentioned in the previous section. To this end, we consider a  system with $N_R=N_U=2$ RRHs and UEs with $N_t=N_r=4$ antennas. The positions of the RRHs and the UEs are fixed\footnote{The positions of RRHs are set as ${\pmb{p}}_{R,1} = [307.50 \,\, 233.18]^T$ and ${\pmb{p}}_{R,2} = [430.3 \,\, 192.64]^T$, where ${\pmb{p}}_{R,i}$ is the position of $i$-th RRH with coordinate origin at the lower left corner, and the positions of UEs as ${\pmb{p}}_{U,1} = [363.7 \,\, 316.66]^T$ and ${\pmb{p}}_{U,2} = [438.17 \,\, 107.09]^T$, where ${\pmb{p}}_{U,j}$ is the position of $j$-th UE.} in the area with side $\delta=500m$ as in Fig. \ref{Chap3_Fig:SimulEnvironment}. In the path loss formula Eq. (\ref{Chap3_PL_coef}), we set the reference distance to $d_0=50m$ and the path loss exponent to $\eta = 3$. Throughout, we assume that each RRH has the same fronthaul capacity $\bar C$, that is $\bar C_j = \bar C$ for $j \in \mathcal{N}_R$. We optimize over the power allocation $(P_p, P_d)$ and we set $T_p = N_t$, which was shown to be optimal in \cite{Hassibi03TIT} for a point-to-point link with no fronthaul limitation.

The effect of an increase of the coherence time on the ergodic achievable sum-rate is investigated in Fig. \ref{Fig:UplinkvsCT} with fronthaul capacity $\bar C$ = 6 bits/s/Hz, and power $\bar P$ = 10dB. As expected from information-theoretic considerations, Fig. \ref{Fig:UplinkvsCT} demonstrates that the alternative approach is advantageous, although most of the gains are accrued by means of adaptive quantization. Moreover, it is observed that the performance of the conventional approach without adaptive quantization approaches that of the alternative approach as the coherence time $T$ increases. This is because, for large coherence time $T$, the fraction of fronthaul capacity devoted to training becomes negligible and hence accurate CSI can be obtained at the BBU.

In Fig. \ref{Fig:UplinkvsFC}, we set the power as $\bar P = 10dB$ and the coherence time as $T=10$, and we plot the ergodic achievable sum-rate versus the fronthaul capacity $\bar C$. The main conclusions are consistent with those discussed above for Fig. \ref{Fig:UplinkvsCT}. Moreover, it is seen that the performance gain of the alternative functional split is relevant as long as $\bar C$ is not too large, in which case the performance is limited by the uplink SNR and not by the limited fronthaul capacity.


\section{Downlink: Where to Perform Channel Encoding and Precoding?} \label{sec:downlink}
In this section, we turn to the downlink and address the issue of whether it is more advantageous to implement channel encoding and precoding at the RRHs rather than at the BBU as in the conventional implementation. Specifically, we compare the following two approaches: 

\begin{itemize}
\item the conventional approach, in which the BBU performs channel coding and precoding and then quantizes and forwards the resulting baseband signals on the fronthaul links to the RRHs; 
\item channel encoding and precoding at the RRHs in which the BBU does not perform precoding but rather forwards separately the information messages of a subset of UEs, along with the quantized precoding matrices to the all RRHs, which then perform channel encoding and precoding. 
\end{itemize}

The conventional approach has been studied under a simplified quasi-static, rather than ergodic, channel model \cite{Simeone09EURADVSP, Park13TSP}, while the alternative functional split was investigated by Park \textit{et al.} \cite{Chae13ICC}. This section is adapted from our earlier paper \cite{Kang14arXiv}, to which we refer for further details and proofs. We also note that we focus here on linear precoding, or beamforming, and separate quantization for each RRH, and that related discussion on non-linear precoding and joint fronthaul quantization can be found in the literature \cite{Park13TSP}. 

We start by detailing the system model in Sec. \ref{Chap4_Sec:SM}. In Sec. \ref{Chap4_Sec:PBC}, we study the conventional approach, while the alternative functional split mentioned above is studied in \ref{Chap4_Sec:PBC_CBP}. In Sec. \ref{Chap4_Sec:Numerical Results}, numerical results are presented.


\subsection{System Model} \label{Chap4_Sec:SM}
\begin{figure}[t]
\centering
\includegraphics[width=13.5cm]{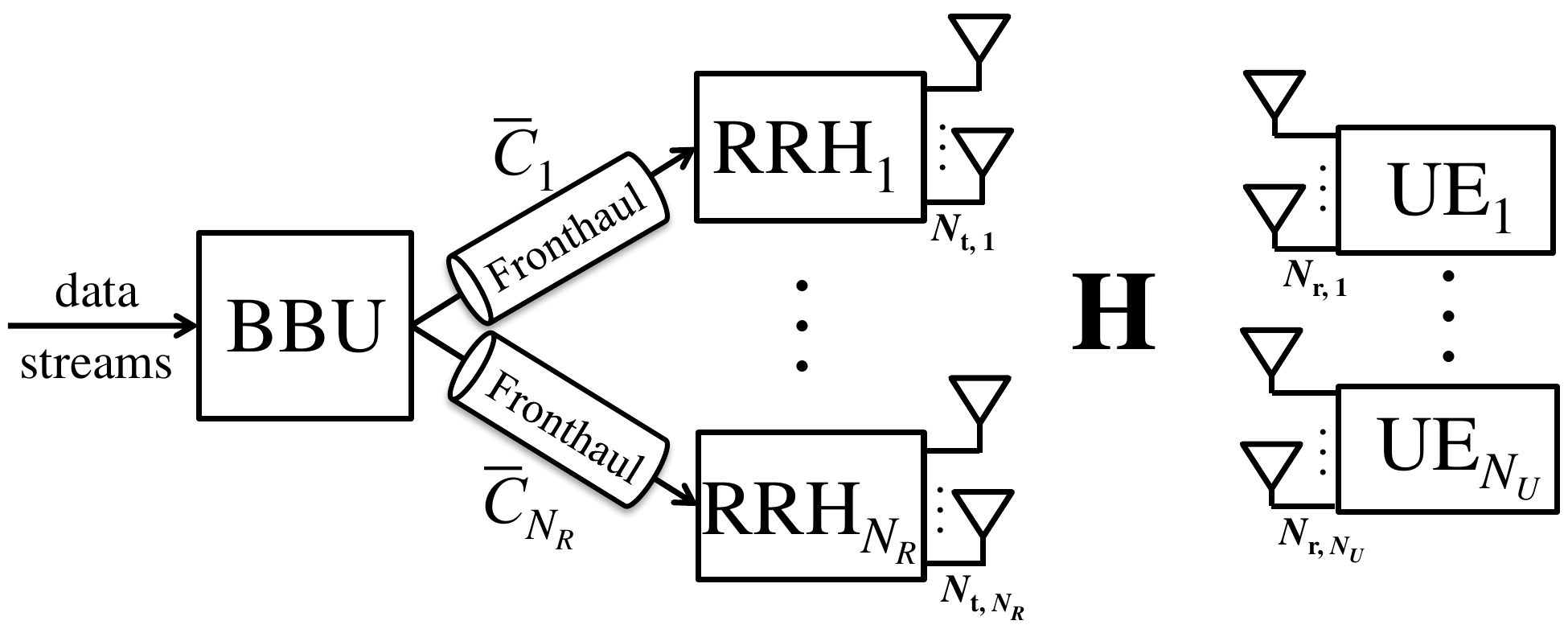}
\caption{Downlink of a C-RAN system consisting of $N_R$ RRHs and $N_U$ UEs. The BBU is connected to each $i$-th RRH with a fronthaul link of capacity $\bar C_i$.}
\label{Chap4_fig:fig1}
\end{figure}
We consider the counterpart downlink C-RAN model of the uplink set-up studied in Sec \ref{sec:uplink}, in which a cluster of $N_R$ RRHs provides wireless service to $N_U$ UEs as illustrated in Fig. \ref{Chap4_fig:fig1}. Most of the baseband processing for all the RRHs in the cluster is carried out at a BBU that is connected to each $i$-th RRH via a fronthaul link of finite capacity $\bar C_i$. Each $i$-th RRH has $N_{t,i}$ transmit antennas and each $j$-th UE has $N_{r,j}$ receive antennas. We denote the set of all RRHs as $\mathcal{N}_R = \{1, \dots, N_R\}$ and the set of all UEs as $\mathcal{N}_U = \{1, \dots, N_U\}$, and we define the number of total transmit antennas as $N_t = \sum_{i=1}^{N_R} N_{t,i}$ and of total receive antennas as $N_r = \sum_{j=1}^{N_U} N_{r,j}$. Moreover, we adopt a block-ergodic channel model in which the fading channels are constant within a coherence period but vary in an ergodic fashion across a large number of coherence periods.

Within each channel coherence period of duration $T$ channel uses, the baseband signal transmitted by the $i$-th RRH is given by a $N_{t,i} \times T$ complex matrix ${\bf{X}}_i$, where each column corresponds to the signal transmitted from the $N_{t,i}$ antennas in a channel use. The $N_{r,j} \times T$ signal ${\bf{Y}}_j$ received by the $j$-th UE in a given channel coherence period, where each column corresponds to the signal received by the $N_{r,j}$ antennas in a channel use, is given by
\begin{equation}
\label{Chap4_RS;UE} {\bf{Y}}_{j} = {\bf{H}}_j {\bf{X}}  + {\bf{Z}}_{j},
\end{equation}
where ${\bf{Z}}_{j}$ is the $N_{r,j} \times T$ noise matrix, which consist of i.i.d. $\mathcal{CN}(0,1)$ entries; ${\bf{H}}_j = [{\bf{H}}_{j1}, \dots, {\bf{H}}_{j_{N_R}}]$ denotes the $N_{r,j} \times N_t$ channel matrix for $j$-th UE, where ${\bf{H}}_{ji}$ is the $N_{r,j} \times N_{t,i}$ channel matrix from the $i$-th RRH to the $j$-th UE; and $ {\bf{X}}$ is the collection of the signals transmitted by all the RRHs, i.e., $ {\bf{X}} = [ {\bf{X}}_{1}^T, \dots,  {\bf{X}}_{N_R}^T]^T$.

We consider the scenario in which the BBU has instantaneous information about the channel matrix ${\bf{H}}$ as well as the case in which the BBU is only aware of the distribution of the channel matrix ${\bf{H}}$, i.e., it has {\textit{stochastic CSI}}. Instead, the UEs always have full CSI about their corresponding channel matrices, as we will state more precisely in the next sections. The transmit signal ${\bf{X}}_i$ has a power constraint given as $T^{-1} E[\|{\bf{X}}_i\|^2] \le \bar P_i$.

While the analysis applies more generally, in order to elaborate on the CSI requirements of the BBU, we consider as a specific channel model of interest the standard Kronecker model, in which the channel matrix ${\bf{H}}_{ji}$ is written as
\begin{equation} \label{Chap4_ChannelMatrix}
{\bf{H}}_{ji} = {\pmb{\Sigma}}_{R, ji}^{1/2} \widetilde {\bf{H}}_{ji} {\pmb{\Sigma}}_{T, ji}^{1/2},
\end{equation}
where the $N_{t,i} \times N_{t,i}$ matrix ${\pmb{\Sigma}}_{T,ji}$ and the $N_{r,j} \times N_{r,j}$ matrix ${\pmb{\Sigma}}_{R,ji}$ are the transmit-side and receiver-side spatial correlation matrices, respectively, and the $N_{r,j} \times N_{t,i}$ random matrix $\widetilde {\bf{H}}_{ji}$ has i.i.d. $\mathcal{CN}(0,1)$ variables and accounts for the small-scale multipath fading \cite{Caire14TIT}. With this model, stochastic CSI entails that the BBU is only aware of the correlation matrices ${\pmb{\Sigma}}_{T,ji}$ and ${\pmb{\Sigma}}_{R,ji}$. Moreover, in case that the RRHs are placed in a higher location than the UEs, one can assume that the receive-side fading is uncorrelated, i.e., ${\pmb{\Sigma}}_{R,ji} = {\bf{I}}_{N_{r,j}}$, while the transmit-side covariance matrix ${\pmb{\Sigma}}_{T,ji}$ is determined by the one-ring scattering model (see \cite{Caire14TIT} and references therein). In particular, if the RRHs are equipped with $\lambda /2$-spaced uniform linear arrays, we have ${\pmb{\Sigma}}_{T,ji} = {\pmb{\Sigma}}_{T}(\theta_{ji}, \Delta_{ji})$ for the $j$-th UE and the $i$-th RRH located at a relative angle of arrival $\theta_{ji}$ and having angular spread $\Delta_{ji}$, where the element $(m,n)$ of matrix ${\pmb{\Sigma}}_{T}(\theta_{ji}, \Delta_{ji})$ is given by
\begin{equation} \label{Chap4_CorrCH}
[{\pmb{\Sigma}}_{T}(\theta_{ji}, \Delta_{ji})]_{m,n} = \frac{\alpha_{ji}}{2 \Delta_{ji}} \int_{\theta_{ji} - \Delta_{ji}}^{\theta_{ji} + \Delta_{ji}} \exp^{-j \pi (m-n) \sin(\phi)} d\phi,
\end{equation}
with the path loss coefficient $\alpha_{ji}$ between the $j$-th UE and the $i$-th RRH being given as Eq. (\ref{Chap3_PL_coef}).

\subsection{Conventional Approach} \label{Chap4_Sec:PBC}
\begin{figure}[t]
\centering
\includegraphics[width=13.5cm]{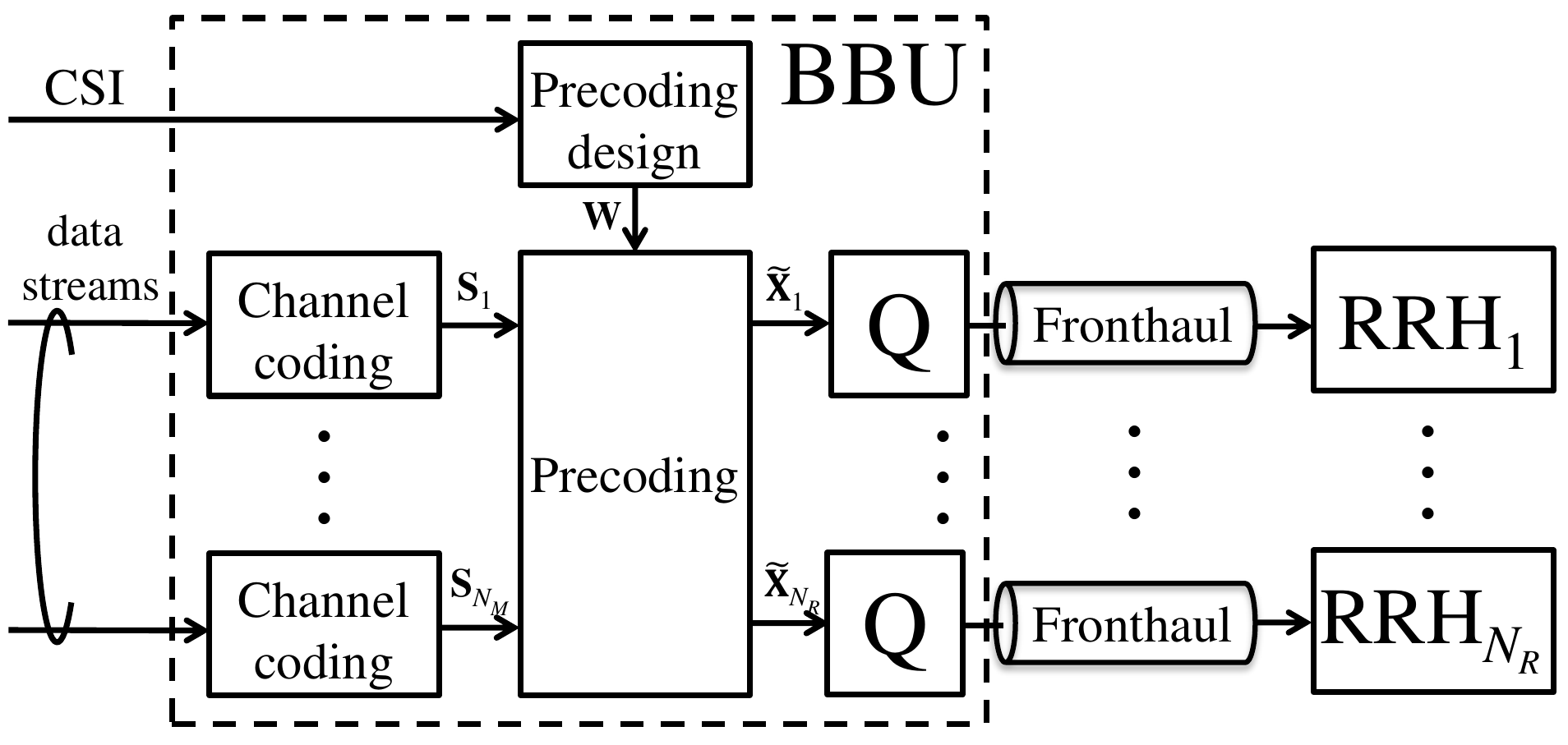}
\caption{Downlink: Conventional approach (``${\text{Q}}$" represents fronthaul compression).}
\label{Chap4_fig:fig2}
\end{figure}
We first describe the conventional approach in Sec. \ref{Chap4_Sec:CAP_PF}. Then, we discuss the joint optimization of fronthaul quantization and precoding with perfect instantaneous channel knowledge at the BBU in Sec. \ref{Chap4_Sec:CAP_PerfectCSI} and under the assumption of stochastic CSI at the BBU in Sec. \ref{Chap4_Sec:CAP_OA}.
\subsubsection{Problem Formulation} \label{Chap4_Sec:CAP_PF}
With the conventional scheme as illustrated in Fig. {\ref{Chap4_fig:fig2}}, the BBU performs channel coding and precoding, and then quantizes the resulting baseband signals so that they can be forwarded on the fronthaul links to the corresponding RRHs. Specifically, channel coding is performed separately for the information stream intended for each UE. This step produces the data signal ${\bf{S}} = [{\bf{S}}_1^\dagger, \dots, {\bf{S}}_{N_U}^\dagger]^\dagger$ for each coherence block, where ${\bf{S}}_j$ is the $M_{j} \times T$ matrix containing, as rows, the $M_j \le N_{r,j}$ encoded data streams for the $j$-th UE. We define the number of total data streams as $M = \sum_{j=1}^{N_U} M_j$ and assume the condition $M \le N_t$. Following standard random coding arguments, we take all the entries of matrix ${\bf{S}}$ to be i.i.d. as $\mathcal{CN}(0,1)$. The encoded data ${\bf{S}}$ is further processed to obtain the transmitted signals ${\bf{X}}$ as detailed below.

The precoded data signal computed by the BBU for any given coherence time can be written as $\widetilde {\bf{X}} = {\bf{W}} {\bf{S}}$, where ${\bf{W}}$ is the $N_t \times M$ precoding matrix. With instantaneous CSI, a different precoding matrix ${\bf{W}}$ is used for different coherence times in the coding block, while, with stochastic CSI, the same precoding matrix ${\bf{W}}$ is used for all coherence times.

In both cases, the precoded data signal $\widetilde {\bf{X}}$ can be divided into the $N_{t,i} \times T$ signals $\widetilde {\bf{X}}_i$ corresponding to $i$-th RRH for all $i \in \mathcal{N}_R$ as $\widetilde {\bf{X}} = [\widetilde {\bf{X}}_1^T, \dots, \widetilde {\bf{X}}_{N_R}^T ]^T$, with $\widetilde {\bf{X}}_i = {\bf{W}}_i^r {\bf{S}}$, where ${\bf{W}}_i^r$ is the $N_{t,i} \times N_r$ precoding matrix for the $i$-th RRH, which is obtained by properly selecting the rows of matrix ${\bf{W}}$ (as indicated by the superscript ``$r$" for ``rows"): the matrix ${\bf{W}}^r_i$ is given as ${\bf{W}}_i^r = {\bf{D}}_i^{r T} {\bf{W}}$, with the $N_t \times N_{t,i}$ matrix ${\bf{D}}_i^r$ having all zero elements except for the rows from $\sum_{k=1}^{i-1} N_{t,k}+1$ to $\sum_{k=1}^i N_{t,k}$, that contain an $N_{t,i} \times N_{t,i}$ identity matrix.

The BBU quantizes each sequence of baseband signal $\widetilde {\bf{X}}_i$ for transmission on the $i$-th fronthaul link to the $i$-th RRH independently. We write the compressed signals ${\bf{X}}_i$ for the $i$-th RRH as
\begin{eqnarray}
\label{Chap4_PDS;EachRRH} {\bf{X}}_i = \widetilde {\bf{X}}_i + {\bf{Q}}_{x,i},
\end{eqnarray}
where the quantization noise matrix ${\bf{Q}}_{x,i}$ is assumed to have i.i.d. $\mathcal{CN}(0, \sigma_{x,i}^2)$ entries. Note that the advantages of joint quantization across multiple RRHs are explored in \cite{Park13TSP} for static channels. Based on Eq. (\ref{Chap4_PDS;EachRRH}), the design of the fronthaul compression reduces to the optimization of the quantization noise variances $\sigma_{x,1}^2, \dots, \sigma_{x,N_R}^2$. The power transmitted by $i$-th RRH  is computed as
\begin{equation} \label{Chap4_PowerConst}
P_i \left({\bf{W}}, \sigma_{x,i}^2 \right) = \frac{1}{T} E  [||{\bf{X}}_i||^2 ] = \textrm{tr} \left( {\bf{D}}_i^{rT} {\bf{W}} {\bf{W}}^\dagger {\bf{D}}_i^r + \sigma_{x,i}^2 {\bf{I}} \right),
\end{equation}
where we have emphasized the dependence of the power $P_i ({\bf{W}}, \sigma_{x,i}^2 )$ on the precoding matrix ${\bf{W}}$ and quantization noise variances $\sigma_{x,i}^2$. Moreover, using standard rate-distortion arguments, the rate required on the fronthaul between the BBU and $i$-th RRH in a given coherence interval can be quantified by $I (\widetilde {\bf{X}}_i; {\bf{X}}_i ) / T$ (see, e.g., \cite[Ch. 3]{GamalBook}), yielding \cite{Kang14arXiv}
\begin{equation} \label{Chap4_BC;CAP}
C_i \left({\bf{W}}, \sigma_{x,i}^2 \right ) =  \log \det \left( {\bf{D}}_i^{rT} {\bf{W}} {\bf{W}}^\dagger {\bf{D}}_i^r+ \sigma_{x,i}^2 {\bf{I}} \right) - N_{t,i} \log \left( \sigma_{x,i}^2 \right),
\end{equation}
so that the fronthaul capacity constraint is $C_i ({\bf{W}}, \sigma_{x,i}^2 ) \le \bar C_i$.

We assume that each $j$-th UE is aware of the effective receive channel matrices $\widetilde {\bf{H}}_{jk} = {\bf{H}}_j {\bf{W}}_k^c$ for all $k \in \mathcal{N}_U$ at all coherence times, where ${\bf{W}}^c_k$ is the $N_t \times N_{r,j}$ precoding matrix corresponding to $k$-th UE, which is obtained from the precoding matrix ${\bf{W}}$ by properly selecting the columns as ${\bf{W}} = [{\bf{W}}^c_1, \dots, {\bf{W}}^c_{N_U}]$. We collect the effective channels in the matrix $\widetilde {\bf{H}}_j = [\widetilde{\bf{H}}_{j1}, \dots, \widetilde{\bf{H}}_{j N_U}] = {\bf{H}}_j {\bf{W}}$. The effective channel $\widetilde {\bf{H}}_j$ can be estimated at the UEs via downlink training.

Under these assumptions, the ergodic achievable rate for the $j$-th UE is computed as $E [R_j^{conv} ({\bf{H}}, {\bf{W}}, {\pmb{\sigma}}_{x}^2 )]$, with $R_j^{conv} ({\bf{H}}, {\bf{W}}, {\pmb{\sigma}}_{x}^2 ) = I_{{\bf{H}}} ({\bf{S}}_j; {\bf{Y}}_j )/T$, where $I_{{\bf{H}}} (\widetilde {\bf{S}}_j; {\bf{Y}}_j )$ represents the mutual information for a fixed realization of the channel matrix ${\bf{H}}$, the expectation is taken with respect to ${\bf{H}}$ and
\begin{eqnarray} \label{Chap4_EAR_CAP}
&& \hspace{-0.7cm} R_j^{conv} \hspace{-0.15cm} \left({\bf{H}}, {\bf{W}}, {\pmb{\sigma}}_{x}^2 \right) \hspace{-0.1cm} =  \log \det \left( \hspace{-0.05cm} {\bf{I}} \hspace{-0.05cm} + \hspace{-0.05cm} {\bf{H}}_j \left( {\bf{W}} {\bf{W}}^\dagger \hspace{-0.1cm} + {\bf{\Omega}}_x \right) {\bf{H}}_j^\dagger \right) \\
\nonumber && \hspace{2.7cm} - \log \det \hspace{-0.1cm} \left ( \hspace{-0.1cm} {\bf{I}} \hspace{-0.05cm} + \hspace{-0.05cm} {\bf{H}}_j \left( \sum_{k \in \mathcal{N}_U \setminus j} \hspace{-0.1cm} {\bf{W}}^c_k {{\bf{W}}_k^c} ^\dagger \hspace{-0.05cm} +\hspace{-0.05cm}  {\bf{\Omega}}_x \right) {\bf{H}}_j^\dagger \right).
\end{eqnarray}
In Eq. (\ref{Chap4_EAR_CAP}), the covariance matrix ${\bf{\Omega}}_x$ is a diagonal with diagonal blocks given as $\textrm{diag} ([\sigma_{x,1}^2 {\bf{I}},$ $\dots, \sigma_{x,N_R}^2 {\bf{I}}])$ and ${\pmb{\sigma}}_{x}^2 = [\sigma_{x,1}^2, \dots, \sigma_{x,N_R}^2]^{T}$.

The ergodic achievable weighted sum-rate can be optimized over the precoding matrix ${\bf{W}}$ and the compression noise variances ${\pmb{\sigma}}_{x}^2$ under fronthaul capacity and power constraints. In the next subsections, we consider separately the cases with instantaneous and stochastic CSI.

\subsubsection{Instantaneous CSI} \label{Chap4_Sec:CAP_PerfectCSI}
In the case of instantaneous channel knowledge at the BBU, the design of the precoding matrix ${\bf{W}}$ and the compression noise variances ${\pmb{\sigma}}_{x}^2$, is adapted to the channel realization ${\bf{H}}$ for each coherence block. To emphasize this fact, we use the notation ${\bf{W}}({\bf{H}})$ and ${\pmb{\sigma}}_{x}^2({\bf{H}})$. The problem of optimizing the ergodic weighted achievable sum-rate with given weights $\mu_j \ge 0$ for $j \in \mathcal{N}_M$ is then formulated as
\begin{subequations} \label{Chap4_P_CAP_wPerfectCSI:WER_STC}
\begin{eqnarray}
\underset { {\bf{W}}({\bf{H}}), {\pmb{\sigma}}_{x}^2({\bf{H}}) }{\textrm{maximize}} && \sum_{j\in \mathcal{N}_U} \mu_j E \left[ R_j^{conv} \left({\bf{H}}, {\bf{W}}({\bf{H}}), {\pmb{\sigma}}_{x}^2({\bf{H}}) \right) \right] \label{Chap4_OF_CAP_wPerfectCSI:WER_STC} \\
\hspace{0.5cm} \textrm{s.t.} \hspace{0.5cm} && C_i \left({\bf{W}}, \sigma_{x,i}^2({\bf{H}}) \right) \le \bar C_i,  \label{Chap4_BC_CAP_wPerfectCSI:WER_STC} \\
&& P_i \left({\bf{W}}({\bf{H}}), \sigma_{x,i}^2({\bf{H}}) \right) \le \bar P_i,  \label{Chap4_PC_CAP_wPerfectCSI:WER_STC}
\end{eqnarray}
\end{subequations}
where Eq. (\ref{Chap4_BC_CAP_wPerfectCSI:WER_STC})-(\ref{Chap4_PC_CAP_wPerfectCSI:WER_STC}) apply for all $i \in \mathcal{N}_R$ and all channel realizations ${\bf{H}}$. Due to the separability of the fronthaul and power constraints across the channel realizations ${\bf{H}}$, the problem in Eq. (\ref{Chap4_P_CAP_wPerfectCSI:WER_STC}) can be solved for each ${\bf{H}}$ independently. Note that the achievable rate in Eq. (\ref{Chap4_OF_CAP_wPerfectCSI:WER_STC}) and the fronthaul constraint in Eq. (\ref{Chap4_BC_CAP_wPerfectCSI:WER_STC}) are non-convex. However, the functions $R_j^{conv} ({\bf{H}}, {\bf{W}}({\bf{H}}), {\pmb{\sigma}}_{x}^2({\bf{H}}) )$ and $C_i ({\bf{W}}({\bf{H}}), \sigma_{x,i}^2({\bf{H}}) )$ are difference of convex (DC) functions of the covariance matrices $\widetilde {\bf{V}}_j({\bf{H}}) = \widetilde{\bf{W}}_j^c({\bf{H}}) \widetilde{\bf{W}}_j^{c \dagger}({\bf{H}})$ for all $j \in \mathcal{N}_U$ and the variance ${\pmb{\sigma}}_{x}^{2}({\bf{H}})$. The resulting rank-relaxed problem can be tackled via the Majorization-Minimization (MM) algorithm as detailed in \cite{Park13TSP}, from which a feasible solution of problem in Eq. (\ref{Chap4_P_CAP_wPerfectCSI:WER_STC}) can be obtained. We refer to \cite{Park13TSP} for details.
\subsubsection{Stochastic CSI} \label{Chap4_Sec:CAP_OA}
With only stochastic CSI at the BBU, in contrast to the case with instantaneous CSI, the same precoding matrix ${\bf{W}}$ and compression noise variances ${\pmb{\sigma}}_{x}^2$ are used for all the coherence blocks. Accordingly, the problem of optimizing the ergodic weighted achievable sum-rate can be reformulated as
\begin{subequations} \label{Chap4_P_CAP_woCSI:WER_STC}
\begin{eqnarray}
\underset {{\bf{W}}, {\pmb{\sigma}}_{x}^2}{\textrm{maximize}} && \sum_{j\in \mathcal{N}_U} \mu_j E \left[ R_j^{conv} \left({\bf{H}}, {\bf{W}}, {\pmb{\sigma}}_{x}^2 \right) \right] \label{Chap4_OF_CAP_woCSI:WER_STC} \\
\hspace{0.5cm} \textrm{s.t.} \hspace{0.5cm} && C_i \left({\bf{W}}, \sigma_{x,i}^2 \right) \le \bar C_i,  \label{Chap4_BC_CAP_woCSI:WER_STC} \\
&& P_i \left({\bf{W}}, \sigma_{x,i}^2 \right) \le \bar P_i,  \label{Chap4_PC_CAP_woCSI:WER_STC}
\end{eqnarray}
\end{subequations}
where Eq. (\ref{Chap4_BC_CAP_woCSI:WER_STC})-(\ref{Chap4_PC_CAP_woCSI:WER_STC}) apply to all $i \in \mathcal{N}_R$. In order to tackle this problem, we adopt the Stochastic Successive Upper-bound Minimization (SSUM) method \cite{SSUM_paper}, whereby, at each step, a stochastic lower bound of the objective function is maximized around the current iterate\footnote{We mention here that an alternative method to attack the problem is the strategy introduced in \cite{Yang13SPAWC}.}. To this end, similar to \cite{Park13TSP}, we can recast the optimization over the covariance matrices ${\bf{V}}_j = {\bf{W}}_j^c {{\bf{W}}_j^c}^\dagger$ for all $j \in \mathcal{N}_U$, instead of the precoding matrices ${\bf{W}}_j^c$ for all $j \in \mathcal{N}_U$. We observe that, with this choice, the objective function is expressed as the average of DC functions, while the constraint in Eq. (\ref{Chap4_BC_CAP_woCSI:WER_STC}) is also a DC function, with respect to the covariance ${\bf{V}} = [{\bf{V}}_1 \dots {\bf{V}}_{N_U}]$ and the quantization noise variances ${\pmb{\sigma}}_{x}^2$. Due to the DC structure, locally tight (stochastic) convex lower bounds can be calculated for objective function in Eq. (\ref{Chap4_OF_CAP_woCSI:WER_STC}) and the constraint in Eq. (\ref{Chap4_BC_CAP_woCSI:WER_STC}) (see, e.g., \cite{MMBook_tutorial}).

\begin{table*}[t]
\caption{Design of Fronthaul Compression and Precoding: Conventional Approach with Stochastic CSI} \label{Chap4_SSUM}
\vspace{-0.5cm}
\rule{18cm}{1pt}
\begin{algorithmic}
\State {\textbf{Initialization}}: Initialize the covariance matrices ${\bf{V}}^{(0)}$ and the quantization noise variances ${\pmb{\sigma}}_{x}^{2 \,\, (0)}$, and set $n=0$.
\State {\textbf{repeat (outer loop)}}
\State \indent $n \gets n+1$
\State \indent Generate a channel matrix realization ${\bf{H}}^{(n)}$ using the available stochastic CSI.
\State \indent {\textbf{Initialization}}: Initialize ${\bf{V}}^{(n,0)} = {\bf{V}}^{(n-1)}$ and ${\pmb{\sigma}}_{x}^{2 \,\, (n,0)} = {\pmb{\sigma}}_{x}^{2 \,\, (n-1)}$, and set $r=0$.
\State \indent {\textbf{repeat (inner loop)}}
\State \indent \indent $r \gets r+1$
\begin{eqnarray} \label{Chap4_ConvexProblem_SSUM_inner}
\nonumber \hspace{1.05cm} \max_{{{\bf{V}}, {\pmb{\sigma}}_{x}^{2}}} && \frac{1}{n} \sum_{l=1}^{n} \sum_{j\in \mathcal{N}_U} \mu_j \widetilde R_j^{conv} \left({\bf{H}}^{(l)}, {\bf{V}}, {\pmb{\sigma}}_{x}^{2} | {\bf{V}}^{(l-1)}, {\pmb{\sigma}}_{x}^{2 \,\, (l-1)} \right) \\
\nonumber \hspace{1.05cm} {\textrm{s.t.}} \hspace{0.1cm} && \widetilde C_i \left({\bf{V}}, {{\sigma}}_{x,i}^{2} | {\bf{V}}^{(n,r-1)}, {{\sigma}}_{x,i}^{2 \,\, (n,r-1)} \right) \le \bar C_i, \\
\nonumber \hspace{1.05cm} && P_i \left ({\bf{V}}, \sigma_{x,i}^{2} \right) \le \bar P_i,  \hspace{0.3cm} \textrm{for all} \,\, i \in \mathcal{N}_R.
\end{eqnarray}
\State \indent \indent Update ${\bf{V}}^{(n,r)} \gets {\bf{V}}$ and ${\pmb{\sigma}}_{x}^{2 \,\, (n,r)} \gets {\pmb{\sigma}}_{x}^{2}$.
\State \indent {\textbf{until}} a convergence criterion is satisfied.
\State \indent Update ${\bf{V}}^{(n)} \gets {\bf{V}}^{(n,r)}$ and ${\pmb{\sigma}}_{x}^{2 \,\, (n)} \gets {\pmb{\sigma}}_{x}^{2 \,\, (n,r)}$.
\State {\textbf{until}} a convergence criterion is satisfied.
\State {\textbf{Solution}}: Calculate the precoding matrix ${\bf{W}}$ from the covariance matrices ${\bf{V}}^{(n)}$ via rank reduction as ${\bf{W}}_j = \gamma_j  {\mathbf{\nu}}_{\textrm{max}}^{(M_j)} ({\bf{V}}_j^{(n)})$ for all $j \in \mathcal{N}_U$, where $\gamma_j$ is obtained by imposing $P_i \left({\bf{W}}, \sigma_{x,i}^2 \right) = \bar P_i$ using Eq. (\ref{Chap4_PowerConst}).
\end{algorithmic}
\rule{18cm}{1pt}
\end{table*}

The algorithm proposed in \cite{Kang14arXiv} is based on SSUM \cite{SSUM_paper} and contains two nested loops. At each outer iteration $n$, a new channel matrix realization ${\bf{H}}^{(n)} = [{\bf{H}}^{T\,\,(n)}_1, \dots, {\bf{H}}^{T\,\,(n)}_{N_U}]$ is drawn based on the availability of stochastic CSI at the BBU. For example, with the model in Eq. (\ref{Chap4_ChannelMatrix}), the channel matrices are generated based on the knowledge of the spatial correlation matrices. Following the SSUM scheme, the outer loop aims at maximizing a stochastic lower bound on the objective function, given as
\begin{equation} \label{Chap4_CAP_OFwithSSUM}
\frac{1}{n} \sum_{l=1}^{n} \widetilde R_j^{conv} \left({\bf{H}}^{(l)}, {\bf{V}}, {\pmb{\sigma}}_{x}^{2} | {\bf{V}}^{(l-1)}, {\pmb{\sigma}}_{x}^{2 \,\, (l-1)} \right),
\end{equation}
where $\widetilde R_j^{conv}({\bf{H}}^{(l)}, {\bf{V}}, {\pmb{\sigma}}_{x}^{2} | {\bf{V}}^{(l-1)}, {\pmb{\sigma}}_{x}^{2 \,\, (l-1)} )$ is a locally tight convex lower bound on $R_j^{conv} ({\bf{H}},$ ${\bf{W}}, {\pmb{\sigma}}_{x}^2 )$ around solution ${\bf{V}}^{(l-1)}$, ${\pmb{\sigma}}_{x}^{2 \,\, (l-1)}$ obtained at the $(l-1)$ the outer iteration when the channel realization is ${\bf{H}}^{(l)}$. This can be calculated as (see, e.g., \cite{SSUM_paper})
\begin{eqnarray} \label{Chap4_linearizedSR_CAP}
\nonumber \hspace{-0.7cm} && \widetilde R_j^{conv} \hspace{-0.05cm} \hspace{-0.05cm} \left ({\bf{H}}^{(l)}\hspace{-0.05cm}, {\bf{V}}, {\pmb{\sigma}}_{x}^{2} | {\bf{V}}^{(l-1)}\hspace{-0.05cm}, {\pmb{\sigma}}_{x}^{2 \,\, (l-1)} \hspace{-0.05cm} \right) \hspace{-0.05cm}\hspace{-0.05cm} \triangleq \hspace{-0.05cm} \log \det \hspace{-0.05cm} \left( \hspace{-0.05cm} {\bf{I}} \hspace{-0.05cm}\hspace{-0.05cm} + \hspace{-0.05cm} {\bf{H}}_j^{(l)} \hspace{-0.05cm} \hspace{-0.05cm} \hspace{-0.05cm} \left( \sum_{k=1}^{N_U} \hspace{-0.05cm} {\bf{V}}_k \hspace{-0.05cm}\hspace{-0.05cm}+\hspace{-0.05cm} {\bf{\Omega}}_x \hspace{-0.05cm} \right) \hspace{-0.05cm}\hspace{-0.05cm} {\bf{H}}_j^{(l)\,\dagger} \hspace{-0.05cm}\hspace{-0.05cm} \right) \\
\hspace{-0.7cm} && \hspace{3cm} - f \left( {\bf{I}} + {\bf{H}}_j^{(l)} \pmb{\Lambda}_j^{(l-1)} {\bf{H}}_j^{(l)\,\,\dagger}, {\bf{I}} + {\bf{H}}_j^{(l)} \pmb{\Lambda}_j {\bf{H}}_j^{(l)\,\,\dagger} \right),
\end{eqnarray}
where $\pmb{\Lambda}_j = \sum_{k=1, k \neq j}^{N_U} {\bf{V}}_k + {\bf{\Omega}}_x$, $\pmb{\Lambda}_j^{(l-1)} = \sum_{k=1, k \neq j}^{N_U}$ ${\bf{V}}_k^{(l-1)} + {\bf{\Omega}}_x$, the covariance matrix ${\bf{\Omega}}_x^{(l)}$ is a diagonal matrix with diagonal blocks given as $\textrm{diag} ([\sigma_{x,1}^{2 \,\, (l)} {\bf{I}}, \dots,$ $\sigma_{x,N_R}^{2 \,\, (l)} {\bf{I}}])$ and the linearized function $f({\bf{A}}, {\bf{B}})$ is obtained from the first-order Taylor expansion of the log det function as
\begin{equation} \label{Chap4_LinearFunc}
f({\bf{A}}, {\bf{B}}) \triangleq \log \det \left( {\bf{A}} \right) + \frac{1}{\textrm{ln} 2} \textrm{tr} \left({\bf{A}}^{-1} \left({\bf{B}} - {\bf{A}} \right) \right).
\end{equation}
Since the maximization of Eq. (\ref{Chap4_CAP_OFwithSSUM}) is subject to the non-convex DC constraint in Eq. (\ref{Chap4_BC_CAP_woCSI:WER_STC}), the inner loop tackles the problem via the MM algorithm i.e., by applying successive locally tight convex lower bounds to the left-hand side of the constraint in Eq. (\ref{Chap4_BC_CAP_woCSI:WER_STC}) \cite{MMBook}. Specifically, given the solution ${\bf{V}}^{(n,r-1)}$ and  ${\pmb{\sigma}}_{x}^{2\,\,(n,r-1)}$ at $(r-1)$-th inner iteration of the $n$-th outer iteration, the fronthaul constraint in Eq. (\ref{Chap4_BC_CAP_woCSI:WER_STC}) at the $r$-th inner iteration can be locally approximated as
\begin{eqnarray} \label{Chap4_linearizedBR_CAP}
&&\hspace{-0.7cm}\widetilde C_i \left( {\bf{V}}, {{\sigma}}_{x,i}^{2} | {\bf{V}}^{(n,r-1)}, {{\sigma}}_{x,i}^{2 \,\, (n,r-1)} \right) \triangleq \\
\nonumber && \hspace{-0.7cm} f \hspace{-0.08cm} \left( \sum_{k=1}^{N_U} \hspace{-0.08cm} {\bf{D}}_i^{rT} {\bf{V}}_k^{(n,r-1)} {\bf{D}}_i^r \hspace{-0.1cm} + \hspace{-0.1cm} \sigma_{x,i}^{2\,\,(n,r-1)} {\bf{I}}, \sum_{k=1}^{N_U}\hspace{-0.08cm} {\bf{D}}_i^{rT} {\bf{V}}_k {\bf{D}}_i^r \hspace{-0.1cm} + \hspace{-0.1cm} \sigma_{x,i}^{2} {\bf{I}} \right) \hspace{-0.1cm} - \hspace{-0.1cm} N_{t,i} \log \hspace{-0.05cm} \left( \sigma_{x,i}^{2}\right)\hspace{-0.1cm}.
\end{eqnarray}
The resulting combination of SSUM and MM for the solution of problem in Eq. (\ref{Chap4_P_CAP_woCSI:WER_STC}) is summarized in Table Algorithm \ref{Chap4_SSUM}. The algorithm is completed by calculating, from the obtained solution ${\bf{V}}^{*}$ of the relaxed problem, the precoding matrix ${\bf{W}}$ by using the standard rank-reduction approach \cite{BoydSemiProg}, which is given as ${\bf{W}}_j^{*} = \gamma_j {\mathbf{\nu}}_{\textrm{max}}^{(M_j)} ({\bf{V}}_j^{*})$ with the normalization factor $\gamma_j$, selected so as to satisfy the power constraint with equality, namely $P_i ({\bf{W}}, \sigma_{x,i}^2 ) = \bar P_i$.

We finally note that, since the approximated functions in Eq. (\ref{Chap4_linearizedSR_CAP}) and Eq. (\ref{Chap4_linearizedBR_CAP}) are local lower bounds, the algorithm provides a feasible solution of the relaxed problem at each inner and outer iteration (see, e.g., \cite{SSUM_paper}).
\subsection{Channel Encoding and Precoding at the RRHs} \label{Chap4_Sec:PBC_CBP}
\begin{figure}[t]
\centering
\includegraphics[width=13.5cm]{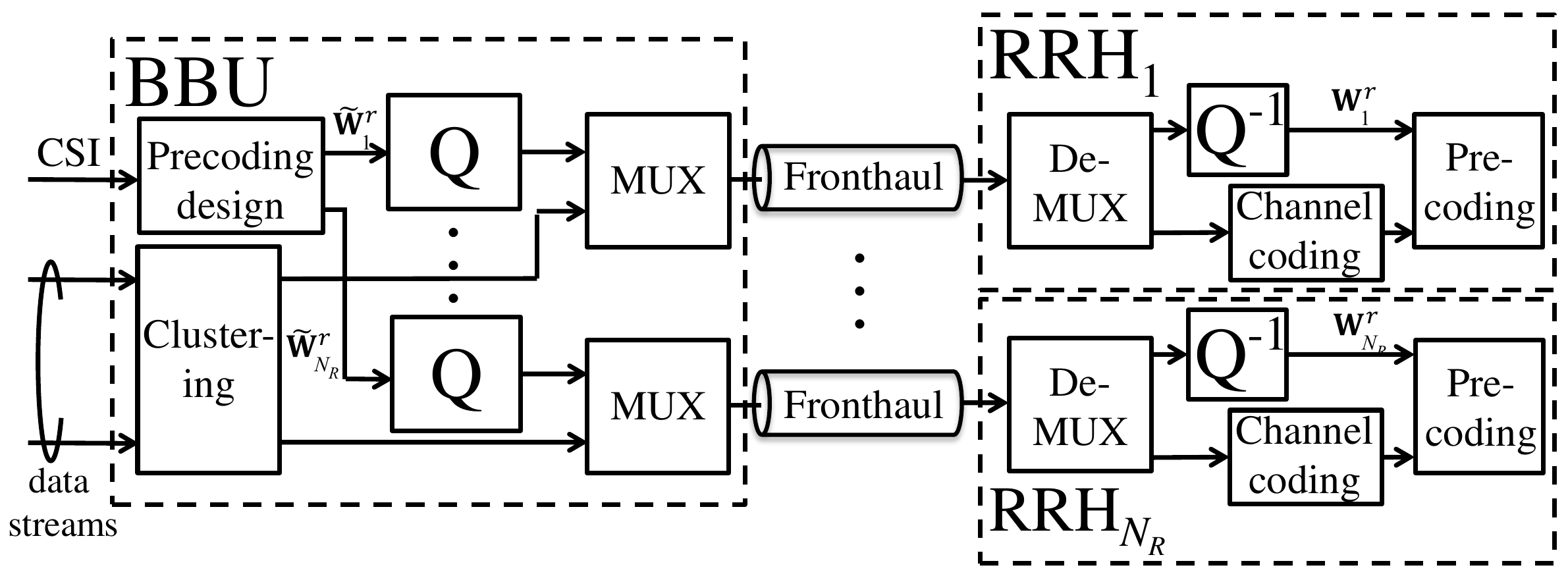}
\caption{Downlink: Alternative functional split (``${\text{Q}}$" and ``${\text{Q}}^{-1}$" represents fronthaul compression and decompression, respectively).}
\label{Chap4_fig:fig3}
\end{figure}
With this alternative functional split, the BBU calculates the precoding matrices, but does not perform precoding. Instead, as illustrated in Fig. {\ref{Chap4_fig:fig3}}, it uses the fronthaul links to communicate the information messages of a given subset of UEs to each RRH, along with the corresponding compressed precoding matrices. Each RRH can then encode and precode the messages of the given UEs based on the information received from the fronthaul link. As it will be discussed, with this approach, a preliminary clustering step is generally advantageous whereby each UE is assigned to a subset of RRHs. In the following, we first describe the strategy in Sec. \ref{Chap4_CBP_PF}. Then we discuss the design problem for fronthaul quantization and precoding under instantaneous CSI in Sec. \ref{Chap4_Sec:CBP_PerfectCSI} and with stochastic CSI in Sec. \ref{Chap4_Sec:CBP_Stochastic}.

\subsubsection{Problem Formulation} \label{Chap4_CBP_PF}
As shown in Fig. \ref{Chap4_fig:fig3}, the precoding matrix $\widetilde {\bf{W}}$ and the information streams are separately transmitted from the BBU to the RRHs, and the received information bits are encoded and precoded at each RRH using the received precoding matrix. Note that, with this scheme, the transmission overhead over the fronthaul depends on the number of UEs supported by a RRH, since the RRHs should receive all the corresponding information streams.

Given the above, we allow for a preliminary clustering step at the BBU whereby each RRH is assigned by a subset of the UEs. We denote the set of UEs assigned by $i$-th RRH as $\mathcal{M}_i \subseteq \mathcal{N}_U$ for all $i \in \mathcal{N}_R$. This implies that $i$-th RRH only needs the information streams intended for the UEs in the set $\mathcal{M}_i$. We also denote the set of RRHs that serve the $j$-th UE, as $\mathcal{B}_j = \{i|j\in \mathcal{M}_i\} \subseteq \mathcal{N}_R$ for all $j \in \mathcal{N}_U$. We use the notation $\mathcal{M}_i[k]$ and $\mathcal{B}_j[m]$ to respectively denote the $k$-th UE and $m$-th RRH in the sets $\mathcal{M}_i$ and $\mathcal{B}_j$, respectively. We define the number of all transmit antennas for the RRHs, which serve the $j$-th UE, as $N_{t, \mathcal{B}_j}$. We assume here that the sets of UEs assigned by $i$-th RRH are given and not subject to optimization (see Sec. \ref{Chap4_Sec:Numerical Results} for further details).

The precoding matrix $\widetilde {\bf{W}}$ is constrained to have zeros in the positions that correspond to RRH-UE pairs such that the UE is not served by the given RRH. This constraint can be represented as
\begin{eqnarray}
\widetilde {\bf{W}} = \left[{\bf{E}}_1^c \widetilde {\bf{W}}_1^c, \dots, {\bf{E}}_{N_U}^c \widetilde {\bf{W}}_{N_U}^c \right],
\end{eqnarray}
where $\widetilde {\bf{W}}_j^c$ is the $N_{t, \mathcal{B}_j} \times N_{r,j}$ precoding matrix intended for $j$-th UE and RRHs in the cluster $\mathcal{B}_j$, and the $N_t \times N_{t, \mathcal{B}_j}$ constant matrix ${\bf{E}}_j^c$ (${\bf{E}}_j^c$ only has either a 0 or 1 entries) defines the association between the RRHs and the UEs as ${\bf{E}}_j^c = [ {\bf{D}}_{\mathcal{B}_j[1]}^c, \dots, {\bf{D}}_{\mathcal{B}_j[|\mathcal{B}_j|]}^c ]$, with the $N_r \times N_{r,j}$ matrix ${\bf{D}}_j^c$ having all zero elements except for the rows from $\sum_{k=1}^{j-1} N_{r,k}+1$ to $\sum_{k=1}^j N_{r,j}$, which contain an $N_{r,j} \times N_{r,j}$ identity matrix.

The sequence of the $N_{t,i} \times N_{r,\mathcal{M}_i}$ precoding matrices $\widetilde {\bf{W}}_i^r$ intended for each $i$-th RRH for all coherence times in the coding block is compressed by the BBU and forwarded over the fronthaul link to the $i$-th RRH. The compressed precoding matrix ${\bf{W}}^r_i$ for $i$-th RRH is given by
\begin{eqnarray}
\label{Chap4_PM;EachRRH} {\bf{W}}_i^r = \widetilde {\bf{W}}_i^r + {\bf{Q}}_{w,i},
\end{eqnarray}
where the $N_{t,i} \times N_{r, \mathcal{M}_i}$ quantization noise matrix ${\bf{Q}}_{w,i}$ is assumed to have zero-mean i.i.d. $\mathcal{CN}(0, \sigma_{w,i}^2 )$ entries and to be independent across the index $i$. Overall, the $N_t \times N_r$ compressed precoding matrix ${\bf{W}}$ for all RRHs is represented as
\begin{eqnarray}
\label{Chap4_PM;AllRRH} {\bf{W}} = \widetilde {\bf{W}} + {\bf{Q}}_{w},
\end{eqnarray}
where ${\bf{W}} = [{\bf{E}}_1^{r \dagger} {\bf{W}}_{w,1}^\dagger, \dots,  {\bf{E}}_{N_R}^{r \dagger} {\bf{W}}_{w,N_R}^\dagger]^\dagger$, $\widetilde {\bf{W}}$ and ${\bf{Q}}_w$ are similarly defined.

Similar to Eq. (\ref{Chap4_EAR_CAP}), an ergodic rate achievable for $j$-th UE can be written as $E[ R_j^{alt} ({\bf{H}}, \widetilde {\bf{W}}, {\pmb{\sigma}}_{w}^2 )]$, where
\begin{eqnarray} \label{Chap4_ARC;CBP}
\nonumber \hspace{-0.7cm} && R_j^{alt} \hspace{-0.1cm} \left ({\bf{H}}, \widetilde {\bf{W}}, {\pmb{\sigma}}_{w}^2 \right) \hspace{-0.1cm} = \hspace{-0.1cm} \frac{1}{T} I_{{\bf{H}}} \left({\bf{S}}_j; {\bf{Y}}_j  \right) \hspace{-0.1cm} = \hspace{-0.1cm} \log \det \left( {\bf{I}} \hspace{-0.1cm} + \hspace{-0.05cm} {\bf{H}}_j \left(\widetilde {\bf{W}} \widetilde {\bf{W}}^\dagger \hspace{-0.1cm} + \hspace{-0.05cm} {\bf{\Omega}}_w \right) {\bf{H}}_j^\dagger \right) \\
\hspace{-0.7cm} && \hspace{2cm} - \log \det \left ( {\bf{I}} + {\bf{H}}_j \left( \sum_{k \in \mathcal{N}_U \setminus j} \widetilde {\bf{W}}_k^c \widetilde {\bf{W}}_k^{c \dagger} + {\bf{\Omega}}_w \right) {\bf{H}}_j^\dagger \right).
\end{eqnarray}

\subsubsection{Instantaneous CSI} \label{Chap4_Sec:CBP_PerfectCSI}
With perfect CSI at the BBU, as discussed in Sec. \ref{Chap4_Sec:CAP_PerfectCSI}, one can adapt the precoding matrix $\widetilde {\bf{W}}({\bf{H}})$, the user rates $\{R_j({\bf{H}})\}$ and the quantization noise variances ${\pmb{\sigma}}_{w}^2({\bf{H}})$ to the current channel realization at each coherence block. The rate required to transmit precoding information on the $i$-th fronthaul in a given channel realizations ${\bf{H}}$ is given by $C_i ({\bf{H}}, \widetilde {\bf{W}}_i^r , \sigma_{w,i}^2 )/T$, with
\begin{eqnarray} \label{Chap4_BC;CBP}
\hspace{-0.7cm} && \frac{1}{T}C_i\left({\bf{H}}, \widetilde {\bf{W}}_i^r , \sigma_{w,i}^2 \right) \hspace{-0.1cm} = \frac{1}{T} I_{\bf{H}} (\widetilde {\bf{W}}_i^r; {\bf{W}}_i^r ) \\
\nonumber \hspace{-0.7cm} && \hspace{3cm} = \frac{1}{T} \hspace{-0.1cm} \left\{ \hspace{-0.05cm} \log \det \hspace{-0.1cm} \left( \hspace{-0.05cm} {\bf{D}}_i^{rT} \widetilde {\bf{W}} \widetilde {\bf{W}}^\dagger {\bf{D}}_i^r \hspace{-0.1cm}+\hspace{-0.05cm} \sigma_{w,i}^2 {\bf{I}}  \right) \hspace{-0.07cm} - \hspace{-0.07cm} N_{t,i} \log \hspace{-0.05cm} \left( \sigma_{w,i}^2 \right) \hspace{-0.05cm} \right\}\hspace{-0.1cm},
\end{eqnarray}
where the rate $C_i(\widetilde {\bf{W}}_i^r, \sigma_{w,i}^2 )$ required on $i$-fronthaul link is defined in Eq. (\ref{Chap4_BC;CAP}). Note that the normalization by $T$ is needed since only a single precoding matrix is needed for each channel coherence interval. Then, under the fronthaul capacity constraint, the remaining fronthaul capacity that can be used to convey precoding information corresponding to the $i$-th RRH is $\bar C_i - \sum_{j \in \mathcal{M}_i} R_j$. As a result, the optimization problem of interest can be formulated as
\begin{subequations} \label{Chap4_P_CBP_wPerfectCSI;WER_LTC}
\begin{eqnarray}
\label{Chap4_OF_CBP_wPerfectCSI;WER_LTC} \underset {\widetilde {\bf{W}}({\bf{H}}), \, {\pmb{\sigma}}^2_{w,i}({\bf{H}}), \{ R_j({\bf{H}})\} }{\textrm{maximize}} \hspace{-1.1cm} && \hspace{0.5cm} \sum_{j\in \mathcal{N}_U} \mu_j R_j({\bf{H}})   \\
\label{Chap4_RC_CBP_wPerfectCSI;WER_LTC} \hspace{0.5cm} s.t. \hspace{0.3cm} && R_j({\bf{H}}) \le R_j^{alt} \left({\bf{H}}, \widetilde {\bf{W}}({\bf{H}}), {\pmb{\sigma}}^2_{w}({\bf{H}}) \right), \\
\label{Chap4_BC_CBP_wPerfectCSI;WER_LTC} \hspace{-1.1cm} && \frac{1}{T} C_i \hspace{-0.05cm} \left(  {\bf{H}}, \widetilde {\bf{W}}_i^r({\bf{H}}) , \sigma_{w,i}^2({\bf{H}}) \right) \le \bar C_i  - \hspace{-0.1cm} \sum_{j \in \mathcal{M}_i} \hspace{-0.15cm} R_j({\bf{H}}), \\
\label{Chap4_PC_CBP_wPerfectCSI;WER_LTC} \hspace{-1.1cm} && P_i \left (\widetilde {\bf{W}}_i^r({\bf{H}}), \sigma_{w,i}^2({\bf{H}}) \right) \le \bar P_i,
\end{eqnarray}
\end{subequations}
where the constraints apply to all channel realization, Eq. (\ref{Chap4_RC_CBP_wPerfectCSI;WER_LTC}) applies to all $j \in \mathcal{N}_U$, Eq. (\ref{Chap4_BC_CBP_wPerfectCSI;WER_LTC}) -  (\ref{Chap4_PC_CBP_wPerfectCSI;WER_LTC}) apply to all $i \in \mathcal{N}_R$ and the transmit power $P_i (\widetilde {\bf{W}}_i^r({\bf{H}}), \sigma_{w,i}^2({\bf{H}}))$ at $i$-th RRH is defined in Eq. (\ref{Chap4_PowerConst}). Similar to Sec. \ref{Chap4_Sec:CAP_PerfectCSI}, the problem in Eq. (\ref{Chap4_P_CBP_wPerfectCSI;WER_LTC}) can be solved for each channel realization ${\bf{H}}$ independently. In addition, each subproblem can be tackled by using MM algorithm \cite{Park13TSP}.

\subsubsection{Stochastic CSI} \label{Chap4_Sec:CBP_Stochastic}
With stochastic CSI at the BBU, the same precoding matrix is used for all the coherence blocks and hence the rate required to convey the precoding matrix $\widetilde {\bf{W}}_i^r$ to each $i$-th RRH becomes negligible. As a result, we can neglect the effect of the quantization noise and set $\sigma_{w,i}^2 = 0$ for all $i \in \mathcal{N}_R$. Accordingly, the fronthaul capacity can be used to transfer the information stream under the constraint $ \sum_{j \in \mathcal{M}_i} R_j \le \bar C_i$, for all $i \in \mathcal{N}_R$. Based on the above considerations, the optimization problem of interest is formulated as
\begin{subequations} \label{Chap4_P_CBP_woCSI;WER_LTC}
\begin{eqnarray}
\label{Chap4_OF_CBP_woCSI;WER_LTC} \hspace{-1cm} \underset {\widetilde {\bf{W}}, \{ R_j\} }{\textrm{maximize}} && \sum_{j\in \mathcal{N}_U} \mu_j R_j   \\
\label{Chap4_RC_CBP_woCSI;WER_LTC} \hspace{0.5cm} s.t. \hspace{0.5cm} && R_j \le E \left [ R_j^{alt} \left({\bf{H}}, \widetilde {\bf{W}}, {\pmb{0}} \right) \right], \\
\label{Chap4_BC_CBP_woCSI;WER_LTC} \hspace{-1cm} && \sum_{j \in \mathcal{M}_i} R_j \le \bar C_i, \\
\label{Chap4_PC_CBP_woCSI;WER_LTC} \hspace{-1cm} && P_i \left (\widetilde {\bf{W}}_i^r, 0 \right) \le \bar P_i,
\end{eqnarray}
\end{subequations}
where Eq. (\ref{Chap4_RC_CBP_woCSI;WER_LTC}) applies to all $j \in \mathcal{N}_U$, Eq. (\ref{Chap4_BC_CBP_woCSI;WER_LTC})-(\ref{Chap4_PC_CBP_woCSI;WER_LTC}) apply to all $i \in \mathcal{N}_R$ and the transmit power $P_i (\widetilde {\bf{W}}_i^r, \sigma_{w,i}^2)$ at $i$-th RRH is defined in Eq. (\ref{Chap4_PowerConst}). In problem Eq. (\ref{Chap4_P_CBP_woCSI;WER_LTC}), the constraint in Eq. (\ref{Chap4_RC_CBP_woCSI;WER_LTC}) is not only non-convex but also stochastic. Similar to Sec. \ref{Chap4_Sec:CAP_OA}, the functions $R_j^{alt}({\bf{H}}, \widetilde {\bf{W}})$ are DC functions of the covariance matrices $\widetilde {\bf{V}}_j = \widetilde{\bf{W}}_j^c \widetilde{\bf{W}}_j^{c \dagger}$ for all $j \in \mathcal{N}_U$, hence opening up the possibility to develop a solution based on SSUM. We refer to \cite{Kang14arXiv} for details on the resulting algorithm.
\subsection{Numerical Results} \label{Chap4_Sec:Numerical Results}
In this section, we compare the performance of the conventional approach and the alternative split. To this end,  we consider RRHs and UEs to be randomly located in a square area with side $\delta=500m$ as in Fig. \ref{Chap3_Fig:SimulEnvironment}. As in Sec. \ref{Sec:Numerical Results uplink}, in the path loss formula Eq. (\ref{Chap3_PL_coef}), we set the reference distance to $d_0=50m$ and the path loss exponent to $\eta = 3$. We assume the spatial correlation model in Eq. (\ref{Chap4_CorrCH}) with the angular spread $\Delta_{ji} = \arctan (r_s/d_{ji})$, with the scattering radius $r_s = 10m$ and with $d_{ji}$ being the Euclidean distance between the $i$-th RRH and the $j$-th UE. Throughout, we consider that the every RRH is subject to the same power constraint $\bar P$ and has the same fronthaul capacity $\bar C$; that is $\bar P_i = \bar  P$ and $\bar C_i = \bar C$ for $i \in \mathcal{N}_R$. Moreover, in the alternative split scheme, the UE-to-RRH assignment is carried out by choosing, for each RRH, the $N_c$ UEs that have the largest instantaneous channel norms for instantaneous CSI and the largest average channel matrix norms for stochastic CSI. Note that this assignment is done for each coherence block in the former case, while in the latter the same assignment holds for all coherence blocks. Note also that a given UE is generally assigned to multiple RRHs.

\begin{figure}[t]
\centering
\includegraphics[width=12cm]{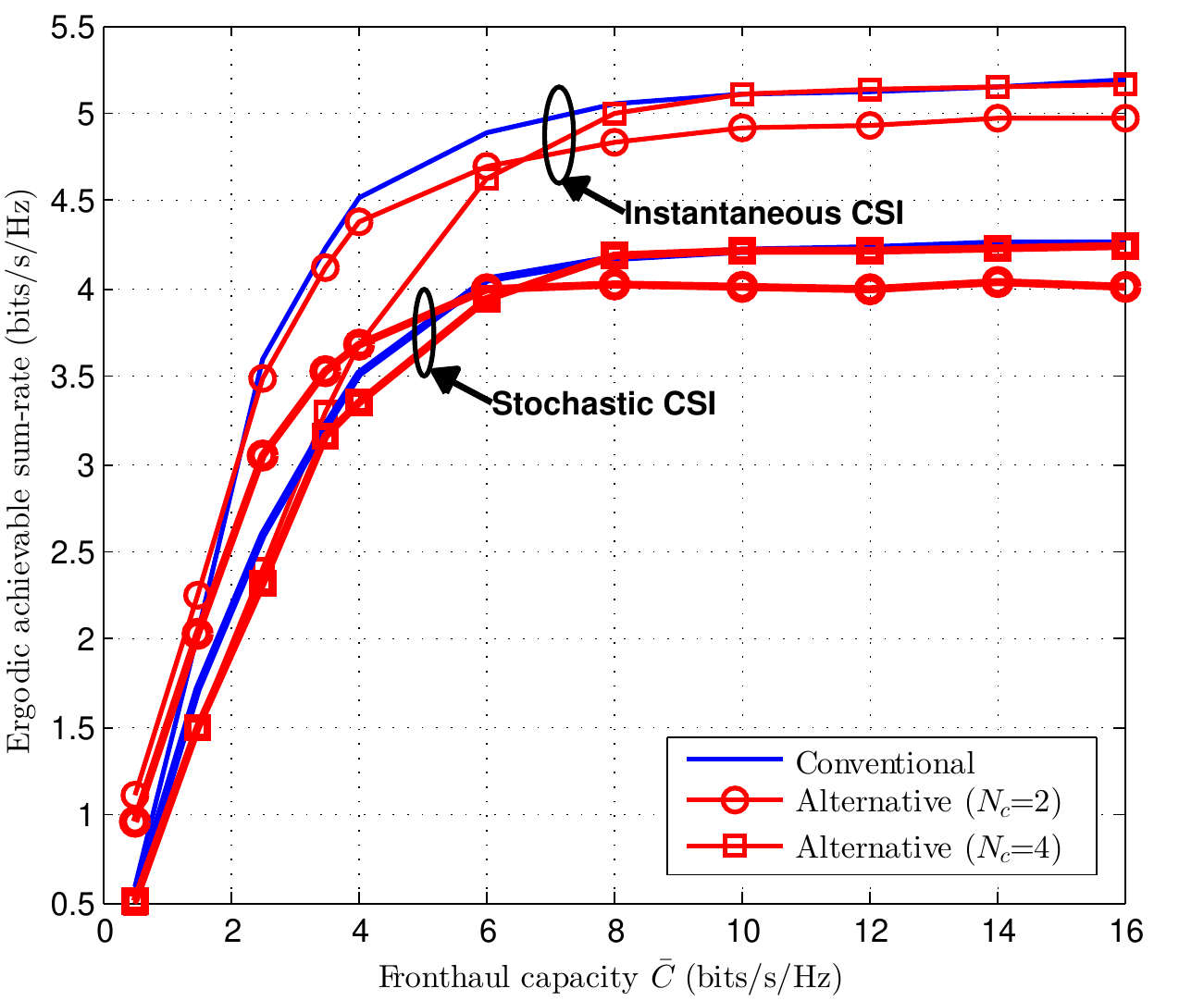}
\caption{Ergodic achievable sum-rate vs. the fronthaul capacity $\bar C$ ($N_R = N_U = 4$, $N_{t,i} = 2$, $N_{r,j}=1$, $\bar P=10$ dB, $T = 20$, and $\mu=1$).}
\label{Chap4_Fig:CAPandCBP_P0T20NR4}
\end{figure}

\begin{figure}[t]
\centering
\includegraphics[width=12cm]{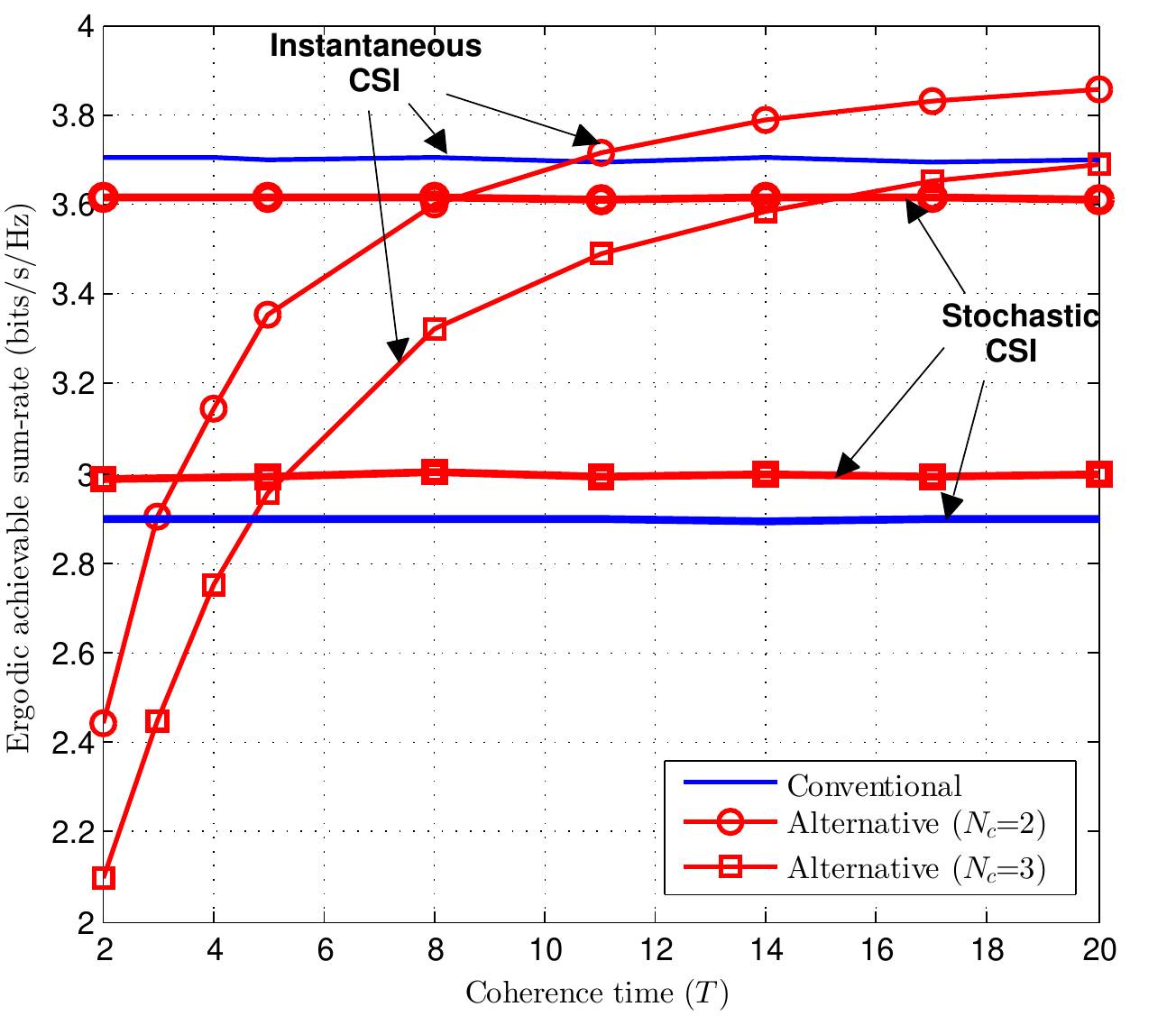}
\caption{Ergodic achievable sum-rate vs. the coherence time $T$ ($N_R = N_U = 4$, $N_{t,i} = 2$, $N_{r,j}=1$, $\bar C = 2$ bits/s/Hz, $\bar P=20dB$, and $\mu=1$).}
\label{Chap4_Fig:CAPandCBP_C8P0NR4}
\end{figure}

The effect of the fronthaul capacity limitations on the ergodic achievable sum-rate is investigated in Fig. \ref{Chap4_Fig:CAPandCBP_P0T20NR4}, where the number of RRHs and UEs is $N_R = N_U = 4$, the number of transmit antennas is $N_{t,i} = 2$ for all $i \in \mathcal{N}_R$, the number of receive antennas is $N_{r,j}=1$ for all $j \in \mathcal{N}_U$, the power is $\bar P=10dB$, and the coherence time is $T = 20$. We first observe that, with instantaneous CSI, the conventional approach strategy is uniformly better than the alternative split as long as the fronthaul capacity is sufficiently large (here $\bar C > 2$). This is due to the enhanced interference mitigation capabilities of the conventional approach resulting from its ability to coordinate all the RRHs via joint baseband processing without requiring the transmission of all messages on all fronthaul links. Note, in fact, that, with the alternative split, only $N_c$ UEs are served by each RRH, and that making $N_c$ larger entails a significant increase in the fronthaul capacity requirements. We will later see that this advantage of the conventional approach is offset by the higher fronthaul efficiency of the alternative split in transmitting precoding information for large coherence periods $T$ (see Fig. \ref{Chap4_Fig:CAPandCBP_C8P0NR4}). Instead, with stochastic CSI, in the low fronthaul capacity regime, here about $\bar C < 6$, the alternative split strategy is generally advantageous due to the additional gain that is accrued by amortizing the precoding overhead over the entire coding block. Another observation is that, for small $\bar C$, the alternative split schemes with progressively smaller $N_c$ have better performance thanks to the reduced fronthaul overhead. Moreover, for large $\bar C$, the performance of the alternative split scheme with $N_c=N_U$, whereby each RRH serves all UEs, approaches that of the conventional scheme.

\begin{figure}[h!]
\centering
\includegraphics[width=12cm]{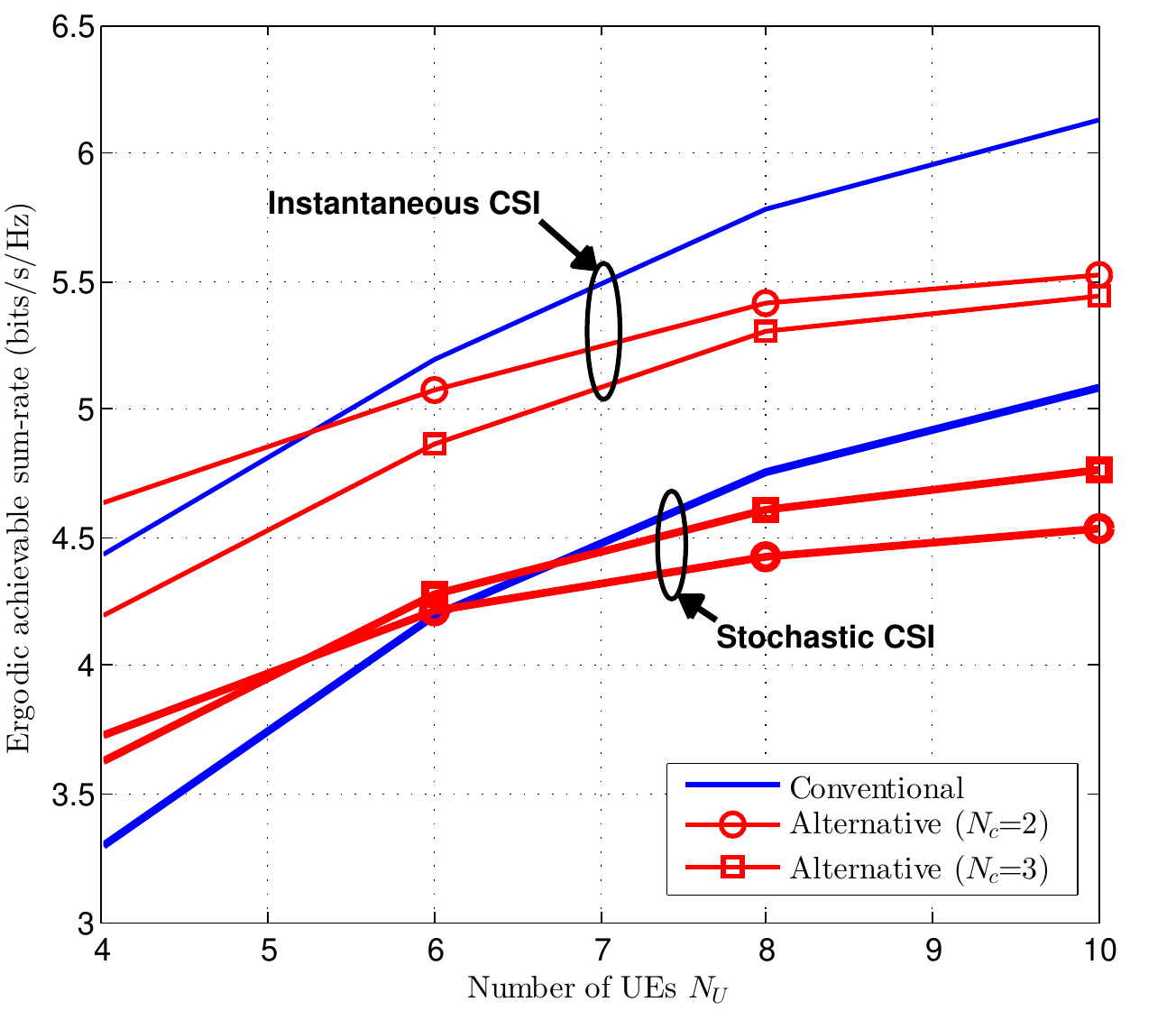}
\caption{Ergodic achievable sum-rate vs. the number of UEs $N_U$ ($N_R = 4$, $N_{t,i} = 2$, $N_{r,j} = 1$, $\bar C=4$ bits/s/Hz, $\bar P=10$ dB, $T = 10$, and $\mu=1$).}
\label{Chap4_Fig:CAPandCBP_C4P10T10_NM}
\end{figure}

Fig. \ref{Chap4_Fig:CAPandCBP_C8P0NR4} shows the ergodic achievable sum-rate as function of the coherence time $T$, with $N_R = N_U = 4$,  $N_{t,i} = 2$, $N_{r,j}=1$, $\bar C = 2$ bits/s/Hz, and $\bar P=20$ dB. As anticipated, with instantaneous CSI, the alternative split is seen to benefit from a larger coherence time $T$, since the fronthaul overhead required to transmit precoding information gets amortized over a larger period. This is in contrast to the conventional approach for which such overhead scales proportionally to the coherence time $T$ and hence the conventional scheme is not affected by the coherence time. As a result, the alternative split can outperform the conventional approach for sufficiently large $T$ in the presence of instantaneous CSI. Instead, with stochastic CSI, the effect is even more pronounced due to the additional advantage that is accrued by amortizing the precoding overhead over the entire coding block.

Finally, in Fig. \ref{Chap4_Fig:CAPandCBP_C4P10T10_NM}, the ergodic achievable sum-rate is plotted versus the number of UEs $N_U$ for $N_R=4$, $N_{t,i} = 2$, $N_{r,j}=1$, $\bar C=4$, $\bar P=10dB$ and $T=10$. It is observed that the enhanced interference mitigation capabilities of the conventional approach without the overhead associated to the transmission of all messages on the fronthaul links yield performance gains for denser C-RANs, i.e., for larger values of $N_U$. This remains true for both instantaneous and stochastic CSI cases.
\section{Concluding Remarks} \label{sec:Conclusion}
In this chapter, we have investigated two important aspects that pertain to the optimal functional split between RRH and BBU at the PHY layer, namely whether uplink channel estimation and downlink encoding/ precoding should be implemented at the RRH or at the BBU. The analysis, based on information-theoretical arguments, and numerical results, built on proposed efficient design algorithms, yields insight into the configurations of network architecture, channel variability and fronthaul capacities in which different functional splits are advantageous. Among the main conclusions, we have argued that the alternative functional split in which uplink channel estimation is performed at the RRH is to be preferred for low or moderate values of the coherence period and fronthaul capacity, and mostly for its capability to enable adaptive quantization based on the channel conditions. Moreover, the alternative functional split in which downlink encoding and precoding are carried out at the RRH is beneficial for lightly loaded networks in the presence of slowly changing channels, particularly under the assumption of stochastic CSI, due to its reduced fronthaul overhead.

We close this chapter with some remark on further related topics and open issues. For the uplink, an aspect that deserves further study is the integration of distributed source coding techniques (or Wyner-Ziv coding) with fronthaul processing for the joint transfer of CSI and data (see \cite{Park13TSP} for some initial discussion). Analogously, for the downlink, the impact of joint, or multivariate, compression, as proposed in \cite{Park13TSP}, on the optimal functional split in the presence of different degrees of CSI at the BBU is an interesting open problem. Finally, the analysis of alternative RRH-BBU functional splits in conjunction with structured coding, or compute-and-forward, techniques calls for further attention (see \cite{Nazer} and references therein). 

\bibliographystyle{IEEEtran}
\bibliography{References}

\end{document}